\documentclass{jaa}
\usepackage{natbib}
\usepackage[normalem]{ulem}
\usepackage{hyperref}
\hypersetup{
    colorlinks=true,
    linkcolor=blue,
    filecolor=magenta,      
    urlcolor=cyan,
    citecolor=blue,
    pdftitle={Overleaf Example},
    pdfpagemode=FullScreen, }
\usepackage{graphicx}
\usepackage{cancel}
\newcommand{\src}{4U 1543--475}
\begin{document}\sloppy
\title{Spectro-temporal investigation of the black hole X-ray transient 4U 1543--475 during the 2021 outburst}

\author{Biki Ram\textsuperscript{1,*}, Manoneeta Chakraborty \textsuperscript{1} and Unnati Kashyap\textsuperscript{2}}
\affilOne{\textsuperscript{1}Department of Astronomy Astrophysics and Space Engineering , IIT Indore, 453552, India.\\}
\affilTwo{\textsuperscript{2}Department of Physics and Astronomy,Texas Tech University, Lubbock, TX 79409-1051, USA}
\twocolumn[{
\maketitle
\corres{phd2101221002@iiti.ac.in}
\msinfo{}{}

\begin{abstract}
We report a detailed spectro-temporal analysis of the black hole (BH) low mass X-ray binary (LMXB) 4U 1543--475 during its 2021 outburst using the data from LAXPC (Large Area X-ray Proportional Counters) and SXT (Soft X-ray Telescope) instrument on board \emph{AstroSat}. We investigate the energy and frequency dependency of the source variability to probe the origin of the disc/coronal fluctuations. Following the state transition (from soft to intermediate state), the emergence of a band-limited noise (BLN) component is observed along with the power law noise (PLN) when the disk is recovering from a sudden decrease in the inner disk radius. A possible correlation between the low-frequency variability amplitude (RMS) and the covering fraction of the non-thermal component is detected. During the final AstroSat observation, a flip-flop phenomenon is reported, where rapid variation in RMS occurs in concurrence with sudden flux transition. 
An indication of the evolution of inner disk temperature along with a significant change in thermal flux was observed 
during the flip-flop phase, arguing for a disk instability-driven origin for this phenomenon. Our results suggest that the long-term variability evolution is primarily affected by the coronal changes, whereas the disk behavior governs the short-term variability evolution.
\end{abstract}

\keywords{Black hole, X-ray binary, accretion,  flip-flop, Individual: 4U 1543--475}}\\]

\section{INTRODUCTION}
Transient black hole (BH) X-ray binary (XRB) sources show frequent outbursts, during which the luminosity of the source increases by orders of magnitude over a period of weeks to months \citep{1997ApJ...491..312C, 2006ARA&A..44...49R}. Following the decay of the outburst, these sources become faint and enter the quiescent phase. The source passes through different states, starting from hard state to hard intermediate state (HIMS) to soft intermediate state (SIMS) to soft state during the outburst \citep{2005Ap&SS.300..107H, 2005A&A...440..207B}. The evolution of the source is displayed by a hardness intensity diagram (HID), which typically follows a `q' shape \citep{2004MNRAS.355.1105F}. The evolution of the XRB between the different spectral states is primarily guided by the change in the mass accretion rate. The thermal emission that dominates in the soft and high-soft state (HSS) is attributed to the optically thick accretion disk, whereas the non-thermal emission that dominates in the hard state is proposed to originate through the comptonization of the disk seed photons by the corona. The emission variability also varies as the source traverses from the hard state (30-40\%) to the soft state ($<$ 5\%) via HIMS and SIMS \citep{2011MNRAS.418.2292M, 2011BASI...39..409B}. In certain cases, the fractional RMS variability in the softer states goes as low as 1\% \citep{2005A&A...440..207B, 2010LNP...794...53B}. During the state transition, noticeable evolution in the timing and spectral properties are observed \citep{2010LNP...794...53B}. The evolution of the sources can sometimes be very complex, as the configuration of the accretion process becomes increasingly ambiguous when the luminosity nears and surpasses the Eddington luminosity. X-ray variability helps to probe the details of the accretion flow and its evolution during an outburst \citep{2008ApJ...675.1407K}. 

The source 4U 1543--475 is a transient BH XRB with an A2V type donor star as its optical counterpart \citep{1984PASJ...36..799K, 1992IAUC.5510....2H, 2004ApJ...610..378P, 2014ApJ...793L..33M}. It was first observed in outburst by the \emph{Uhuru} satellite in 1971 when the intensity was estimated to be 2 Crab \citep{1972ApJ...174L..53M}. Then it went into subsequent outbursts in 1983 \citep{1984PASJ...36..799K}, 1992 \citep{1992IAUC.5510....2H}, 2002 \citep{2004ApJ...610..378P}, and 2021 \citep{2023MNRAS.520.4889P}. The mass function ($f(M)=0.22\pm0.02$), distance (9.1$\pm$ 0.1 kpc), and mass of the black hole were estimated by \citet{1998ApJ...499..375O}. \citet{1997xrb..book.....L} found an ultra-soft component and hard power-law tail in the spectra during its 1992 outburst. The source was observed by \emph{XMM-Newton} in 2002 when \cite{2002ATel..103....1K} found a rapid change in variability from a few percent to tens of percent in less than a day, which indicates a state transition. During the 2002 outburst, the peak luminosity reached 4.2 Crab in the 2-12 keV energy range, where the source was observed to be in a thermally dominant (TD) state. Near the outburst peak, radio flares were observed. As the decay phase of the outburst began, the source entered the steep power law (SPL) state \citep{2004ApJ...610..378P}. 
Recently, \cite{2023MNRAS.520.4889P} performed a spectral analysis of the source and found a dynamic absorption feature between 8-11 keV throughout the 2021 outburst, which was attributed to a fast-moving ionized wind in the disk.  \cite{2023arXiv230708973Y} examined the spectral characteristics of this source using data from \emph{Insight-HXMT (Hard X-ray Modulation Telescope)} and \emph{NICER (Neutron star Interior Composition Explorer)} during the recent outburst and constrained the spin parameter of the source to $a=0.67_{-0.24}^{+0.14}$, assuming a lower viscosity parameter of the disk  ($\alpha =0.01)$. \cite{2023MNRAS.tmp.2157H} conducted a broadband spectral analysis using three different spectral models to explain the emission behavior of the source using a disk continuum with relativistic and non-relativistic models during the 2021 outburst. All the previous studies related to the 2021 outburst of \src \ aim to improve the understanding of radiative behavior and its evolution in high luminosity regimes using spectral analysis. Till now, the timing analysis of this particular outburst has not been performed, which helped us to identify the fluctuation properties of the source in different states.
\\
In spite of the outbursts of low-mass X-ray binaries (LMXB) being observed for decades, there still remain unanswered questions regarding the geometry and evolution of the emission components, especially that of the corona. Multiple theoretical models have been proposed to describe the emission processes in an LMXB and the geometry of the disk and the corona. Furthermore, along with the steady, systematic variation of flux during the outburst with evolving mass accretion rate, certain abrupt, repetitive flux fluctuations have been observed from LMXBs. This behavior is termed as the flip-flop phenomenon and remains yet to be fully understood. These flip-flops show a top-hat-like shape in the light curve. Additionally, flip-flops are observed in coordination with variation in the power density spectrum (PDS) and are believed to be linked to specific state transitions \citep{2020A&A...641A.101B,1991ApJ...383..784M}. Certain flip-flop events are observed either at the peak of the outburst or in proximity to it, \citep{2001ApJS..132..377H,2004A&A...426..587C} and some flip-flops occur during the later phases of the outburst when the flux is declining \citep{2011ApJ...731L...2K,2012A&A...541A...6S}. 
\cite{2007A&ARv..15....1D} mentioned a fundamental framework describing how the interplay between irradiation-controlled hydrogen ionization instability and radiation pressure instability can play a vital role in flux variations over different timescales.  This flip-flop phenomenon has been reported in selective black hole X-ray binary systems, namely, Swift J1658.2-4242 \citep{2020A&A...641A.101B}, MAXI J1348–630 \citep{1994ApJ...435..398M, 2021MNRAS.505..713J}, GS 1124-683 \citep{1997ApJ...489..272T}, XTE J1550-564 \citep{2001ApJS..132..377H, 2016ApJ...823...67S}, GRS 1915+105 \citep{2008MNRAS.383.1089S},  MAXI J1535-571 \citep{2018ApJ...866..122H}, XTE J1859+226 \citep{2004A&A...426..587C, 2013ApJ...775...28S}, H1743-322 \citep{2005Ap&SS.300..107H}, GX 339-4 \citep{2003A&A...412..235N}, XTE J1817-330 \citep{2012A&A...541A...6S}, and MAXI J1659-152 \citep{2011ApJ...731L...2K}. The time between adjacent flip-flop flux transitions can exhibit a broad range, from mere fractions of a second to durations exceeding several ks \citep{2001ApJS..132..377H,1991ApJ...383..784M,1997ApJ...489..272T}. Significant deviations in the spectral parameters between different flux levels of the flip-flop phenomenon were reported by \citet{2020A&A...641A.101B}. \cite{1991ApJ...383..784M} observed a modest change in the disk blackbody component of the spectra when the flux transitioned from a low level to a high level. However, the corresponding comptonized blackbody component increased by approximately 30\%. A comprehensive and consistent physical interpretation of this phenomenon remains elusive till date.\\
In this work, we focus towards understanding the variability in the emission from \src \ at the high accretion rates using high time-resolution data. This is the first time the variability characteristics and their energy dependence (\S~\ref{sec:3.2}) were examined of the 2021 outburst of this source using \emph{AstroSat} data. We also explained the spectral behavior of the source in Section \ref{sec:3.3}. We discuss in detail the connection between spectral parameters and the variability properties. We have also detected atypical flux variation over short timescales (flip-flop) for the very first time for this source using \emph{AstroSat} data which helped us to understand the insights of the physical mechanism that governed this kind of rapid variations. The corresponding spectro-temporal behavior is reported in Section \ref{sec:3.4}. Finally, the significance of our results is discussed in Section \ref{sec:4}

\begin{table*}
\begin{center}
\scalebox{0.9}{%
    \begin{tabular}{ccccccc}
    \hline
   obs No. & obs ID & Start date  & Start time & End date & End time  &  Exposure time (s)  \\
   &  & (dd-mm-yyyy)& (hh:mm:ss) &(dd-mm-yyyy)&  (hh:mm:ss) &  \\
   \hline
   1& 9000004494 (offset of 40$'$ ) &01-07-2021 & 00:59:28.97 & 01-07-2021&11:53:10.72 & 39214.10 \\             
                 2&9000004526 (offset of 40$'$) &10-07-2021 & 07:23:59.61 &10-07-2021&  18:16:45.77  & 36628.55
                 \\
              3 &9000004552 (offset of 40$'$)& 18-07-2021 & 15:03:39.82 &19-07-2021& 05:20:46.57 &    51422.77 \\
              
              4 &9000004588 &26-07-2021 & 03:05:36.37 &26-07-2021& 08:58:13.78 & 17868.27 \\
              5 &9000004622 &04-08-2021& 09:47:19.21&05-08-2021& 04:59:16.61 & 69111.04\\
             6 & 9000004650& 21-08-2021 & 08:29:35.69& 22-08-2021&02:56:14.74& 66399.05\\
              7 &9000004680 &31-08-2021& 00:12:51.60 &01-09-2021& 09:10:40.26& 117886.78 \\
              8 &9000004686 &04-09-2021 &00:14:44.83 &04-09-2021& 15:35:27.01 &  55237.83\\
              9&9000004704 &23-09-2021 &20:02:12.98 & 26-09-2021&18:31:36.00& 253763.00\\\hline
            \end{tabular}}\\
\caption{Details of \emph{AstroSat} observations over nine different epochs during the decay phase of the 2021 outburst of 4U 1543--475. \label{Tab:1}}
\end{center}
\end{table*}
\section{OBSERVATION AND DATA REDUCTION \label{sec:2}}
In this work, we used nine \emph{AstroSat} observations. The details of the observations are given in Table~\ref{Tab:1}. The LAXPC (Large Area X-Ray proportional counter) detectors are the three identical proportional counters (LAXPC10, LAXPC20, LAXPCX30) onboard \emph{AstroSat}, which cover the energy range of 3-80 keV. The timing resolution of the LAXPC is 10  $\mu$s, which makes it ideal for studying the fast timing properties of X-ray binaries \citep{2017CSci..113..591Y, 2017JApA...38...30A}. The data obtained by LAXPC20 is used for this study as it provides the maximum gain and sensitivity, unlike LAXPC10 and LAXPC30. The total effective area of LAXPC20 is $\sim$6000 $cm^2$ around 15 keV, and the dead time is  $\sim42$  $\mu$s. The field of view of the detector is 1$^{\circ}$ $\times$ 1 $^{\circ}$ and energy resolution is 15-20  $\%$ at around 30 keV. During the first three epochs, due to the extreme brightness of the source, the LAXPC observations were taken offset/off-axis. For this work, we utilized the event mode data (at a time resolution of 10 $\mu$s) for all nine epochs of the source. The level 2 data were extracted for further analysis using the {\tt LaxpcSoft} software \citep{2017ApJS..231...10A}. The software runs on the level 1 data, filters for good time intervals (GTI), and takes care of the satellite and instrumental effects. Further, the standard products, including light curve, spectra, and background files, were generated using data from all the anodes. Barycenter correction was applied to the LAXPC level 2 data using the {\tt as1bary} tool.
\begin{figure}
\centering
\includegraphics[scale=0.6]{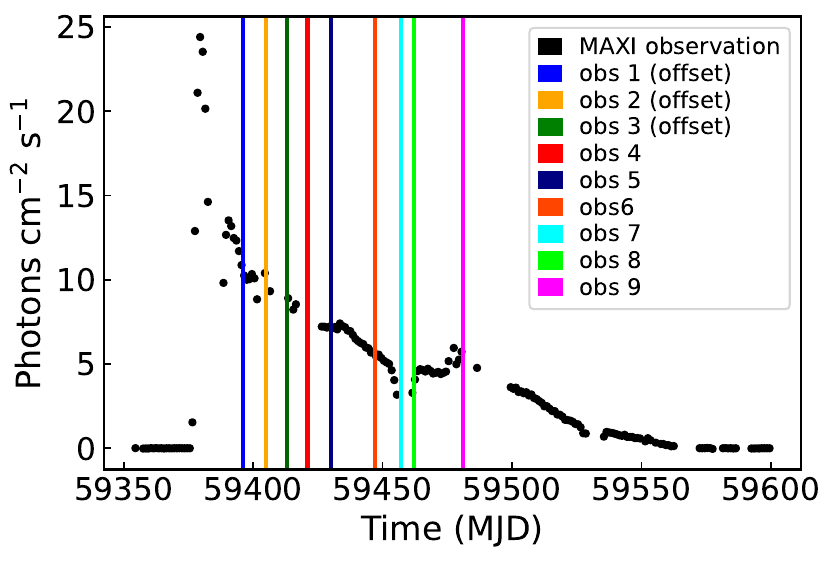}
\caption{1 day averaged light curve of 4U 1543-475 using MAXI (2-20 keV) in black color, the nine LAXPC observations are marked using vertical lines of different colors. \label{Fig:1} }
\centering
\includegraphics[scale=0.5]{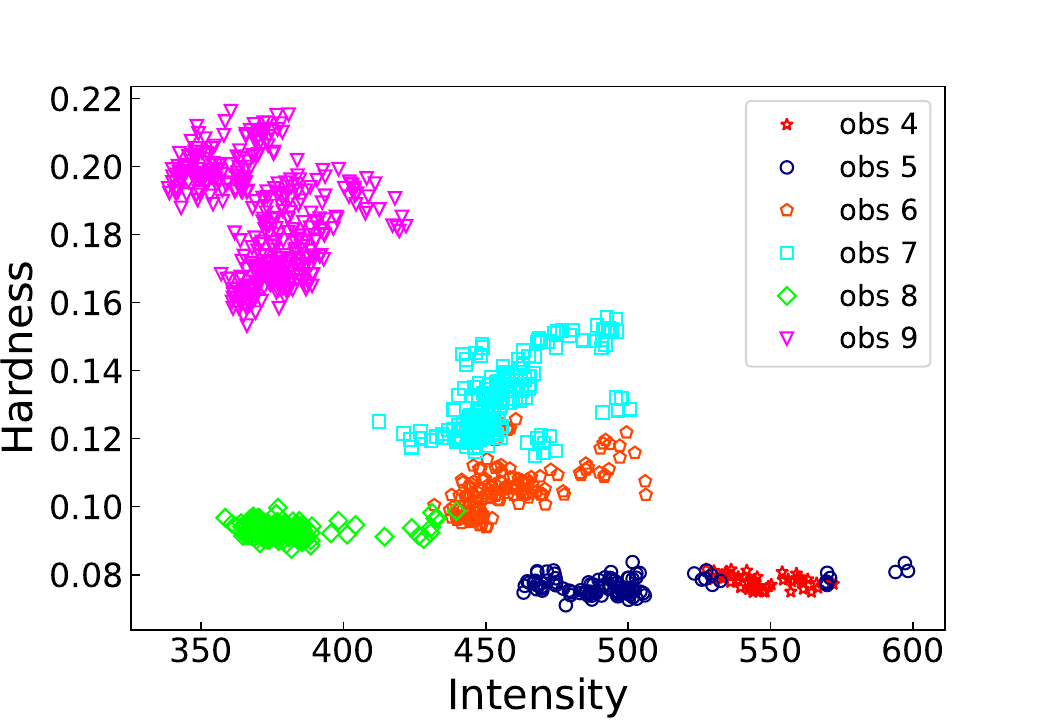}
\caption{The hardness intensity diagram (HID) of LAXPC observations of 4U 1543--475 during the 2021 outburst for the last \textbf{six} on-axis observations. 
The hardness here is defined as the count rate ratio of hard photons (8.46--18 keV) and soft photons (5.14--8.46 keV). Each point corresponds to 128 s bin size. \label{Fig:2}}
\end{figure}

SXT is the Soft X-ray Telescope onboard \emph{AstroSat}, which covers 0.3-8 keV energy range \citep{2017JApA...38...29S}. The effective area of this instrument is 90 $cm^2$ at 1.5 keV. 
We have used the PC mode level2 SXT data, which corresponds to a time resolution of 2.3775 s. We used the {\tt SXTEVTMERGERTOOL} to merge the cleaned event files for the different orbits for each observation.  As the source was exhibiting an extremely bright emission, we saw a pile-up effect in the PC mode data of the on-axis observations. Thus, to remove the effect of pile-up, we excluded a large portion of the central CCD. We selected an annular region of $12-18'$ \citep{2023MNRAS.tmp.2157H} for the on-axis observations and a circular region of $18'$ for the offset observations to extract the final products. We chose grades 0-12 for the offset observation and single-pixel events by filtering the grade 0 events for the on-axis events as the count rate was too high for the on-axis observations. We used {\tt XSELECT} version 2.4m to extract the images, light curves, and spectra. The modified {\tt SXTARFModule} v03 was employed to extract the modified ancillary response using the default ancillary response file of SXT (sxt\_pc\_excl00\_v04\_20190608.arf).  We used the response files sxt\_pc\_mat\_g0to12.rmf for the offset observations and sxt\_pc\_mat\_g0.rmf for the on-axis observations, and the background sky spectrum file SkyBkg\_comb\_EL3p5\_Cl\_Rd16p0\_v01.pha  provided by the SXT team during our spectral analysis.

\section{ANALYSIS AND RESULTS }
\subsection{\textbf{Light Curve and HID  \label{sec:3.1}}}
Figure~\ref{Fig:1} represents the \emph{MAXI (Monitor of All-sky X-ray Image)} long-term light curve of \src \ with vertical colored lines marking the epochs of the \emph{AstroSat} observations during its 2021 outburst. A hardness-intensity diagram (HID) was constructed to estimate the spectral state evolution of the source, considering the on-axis observations. Hardness is the ratio between count rates in the hard and soft energy ranges. The offset observations were neglected, as the source count rate was not accurately measured. To calculate the hardness, we used the energy ranges of 5.14--8.46 keV (soft) and 8.46--18 keV (hard) \citep{2019MNRAS.485.3696C}. The consequent hardness values are plotted across the intensity (in the 5.14--18 keV energy band) in Figure~\ref{Fig:2}. Each data point corresponds to a 128 s bin size where different colors/markers represent the different observations, as mentioned in the inset. The hard color values show a stable behavior during the fourth and fifth epochs. A sudden increase of hardness during the sixth observation was observed, followed by another increase in hardness in the seventh observation. Subsequently, the source became comparatively softer for the eighth observation (represented by the light green points) and finally exhibited the hardest emission in the ninth observation.
\vspace{1.0mm}
\subsection{\textbf{Variability Analysis} \label{sec:3.2}}
\begin{figure*}
    \centering
        \includegraphics[scale=0.35]{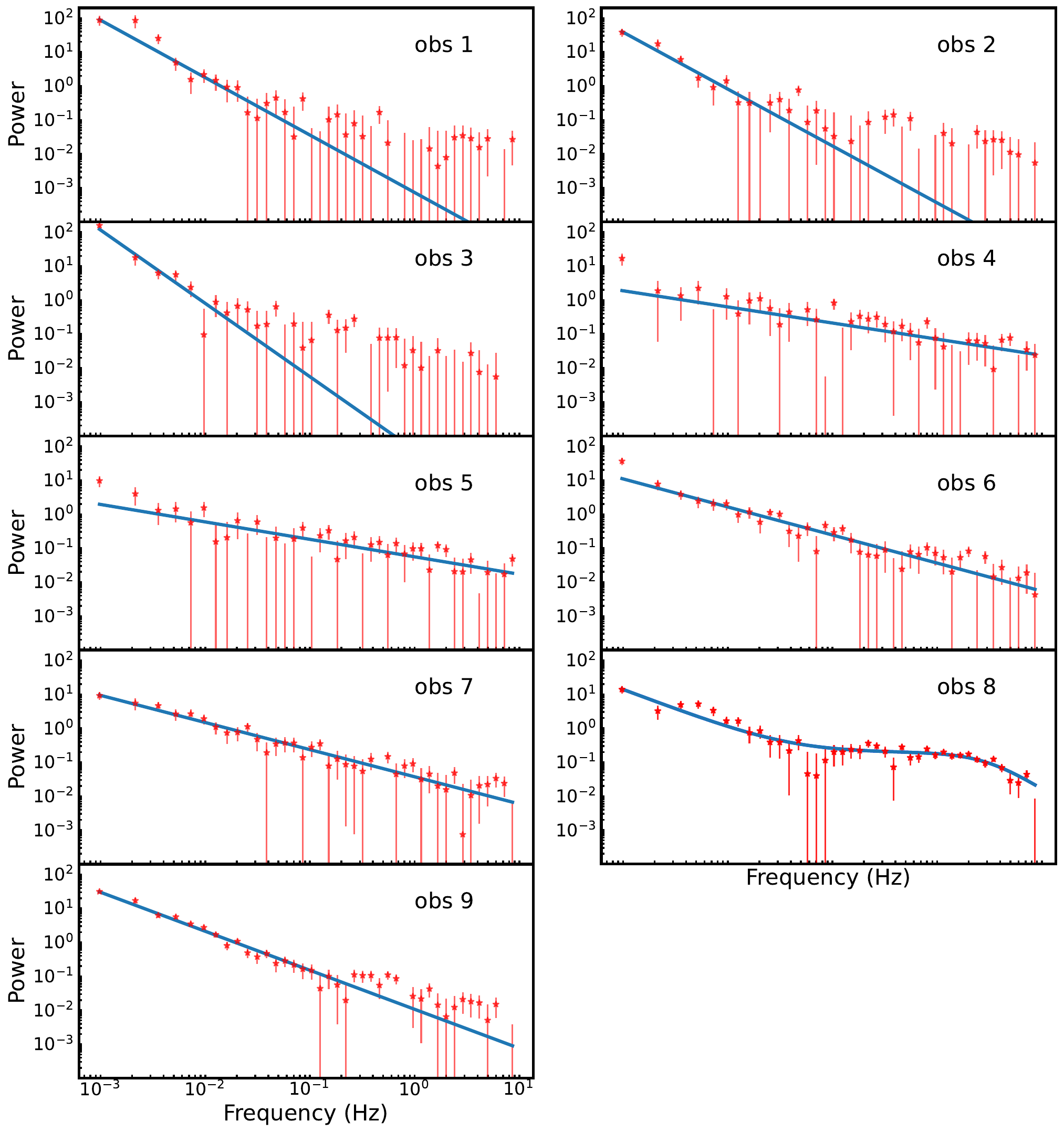}
    \caption{Poisson noise level subtracted, logarithmic re-binned power spectra of 4U 1543--475 using the entire 4-80 keV energy band for all nine LAXPC observations. Considering an F-test threshold criterion of $3 \sigma$, a power law + zero-centered Lorentzian model was required only for the eighth observation, whereas the power law model provided the best fit for the others. The corresponding fits are shown by the solid blue line, and the different panels represent the power spectra for the different observations. \label{Fig:3} }
\end{figure*}

The light curves were extracted from level 2 event mode data for all LAXPC observations, and a power spectral analysis was employed to investigate the variability of the source. For each 1024 s segment of the light curve, Leahy normalized power spectra were constructed \citep{1983ApJ...266..160L}. A Nyquist frequency of 128 Hz was used for the power spectral analysis. We have used the energy range of 4-80 keV for the variability analysis. The time average of power spectra has been computed to reduce noise, as recommended by \cite{1989ASIC..262...27V}. Notably, low-frequency red noise was consistently observed in the power spectra across all eight observations. A logarithmic re-binning procedure was applied to improve the signal-to-noise ratio (SNR) of the power spectra, utilizing a binning factor of 0.2. An instrumental feature around 51 Hz was observed in the power spectra only for the on-axis observations, {possibly detectable} due to the { higher observed} count rate. We have subtracted the Poisson noise by averaging power above 10 Hz while excluding the instrumental feature (45-55 Hz). Figure~\ref{Fig:3} displays the Poisson power subtracted rebinned power spectra for all nine LAXPC observations. The power spectra for the first seven and the ninth observations are well described by the power law model (M1).
\[M1= a  \times (f/0.001)^{-\alpha}\] where $a$ and $\alpha$ are the model parameters and $f$ is the frequency of variability.
However, an additional component was necessary for the eighth observation to accurately model the band-limited noise (BLN) detected/present in the power spectra.
To address this, a zero-centered Lorentzian component has been introduced to fit the BLN, resulting in a notable improvement in the fit quality (decrease in reduced $\chi^2$ from 3.02 (d.o.f.=39) to 1.32 (d.o.f.=37)) for the eighth observation. In order to investigate whether the Lorentzian component was also required in the other observations, all power spectra were fitted using a power law + { zero-centered} Lorentzian (M2) model. 

\begin{figure*}
    \centering
        \includegraphics[scale=0.5]{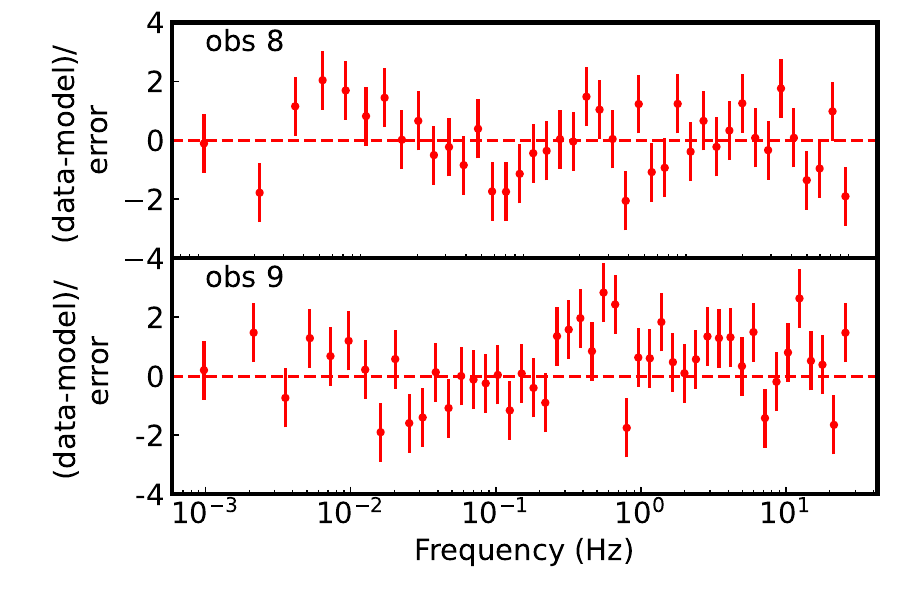}
    \caption{Error-weighted residuals of the Poisson noise subtracted power spectra  fitting using power law+ zero-centred lorentzian model for obs 8 and only power law model for obs 9. \label{Fig:4}}
\end{figure*}
\[M2=  a  \left(\frac{f}{0.001}\right)^{-\alpha}  + \frac{A^2}{\pi}  \frac{w/2} {(w/2)^2 + f^2}\] where $A$ is the amplitude and $w$ is the width of the Lorentzian component. We have conducted an F-test using the chi-squared values of the M1 and M2 models to check the statistical significance of the best-fitting model. A threshold of $3 \sigma$ has been used for rejecting the null hypothesis in the F-test. Therefore, only for the eighth observation, the additional zero-centered Lorentzian component has been considered when the chance probability of the fit improvement was less than 0.27 \% while the M1 model emerged as the best-fit model for the other eight observations. Figure~\ref{Fig:4} represents the error-weighted residuals for the eighth and ninth observations for the corresponding best-fit models (M2 for obs 8 and M1 for obs 9). The residuals support the significant presence of the BLN for the eighth observation. The bottom panel of this plot shows some positive residuals at high frequencies, possibly indicating a very weak presence of BLN in the ninth observation as well; however, its significant presence could not be claimed using the F-test.
The width and the amplitude of the zero-centered Lorentzian for the eighth observation were found to be 6.03 $\pm$ 0.90  and 1.35 $\pm$ 0.07, respectively.

\subsubsection{\textbf{RMS Evolution:} \label{subsec:3.2.1}}
\begin{figure*}
    \centering
\includegraphics[width=1.1\textwidth]{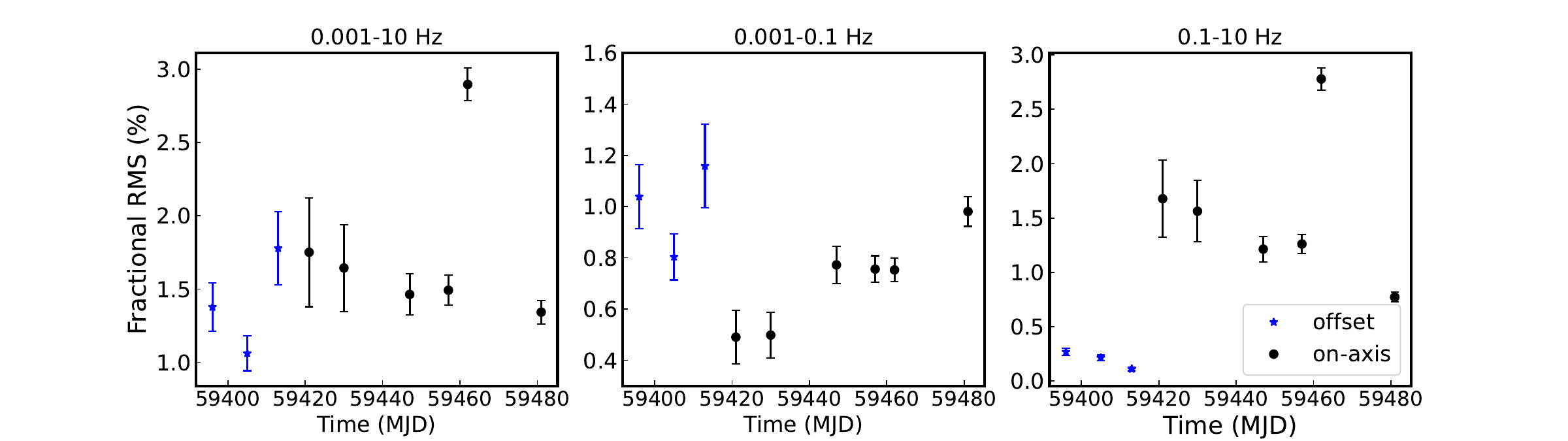}
    \caption{Evolution of fractional RMS amplitude for both offset and on-axis observations in different frequency ranges 0.001-10 Hz (left), 0.001-0.1 Hz (middle) and 0.1-10 Hz (right). \label{Fig:5} }

\includegraphics[scale=0.5]{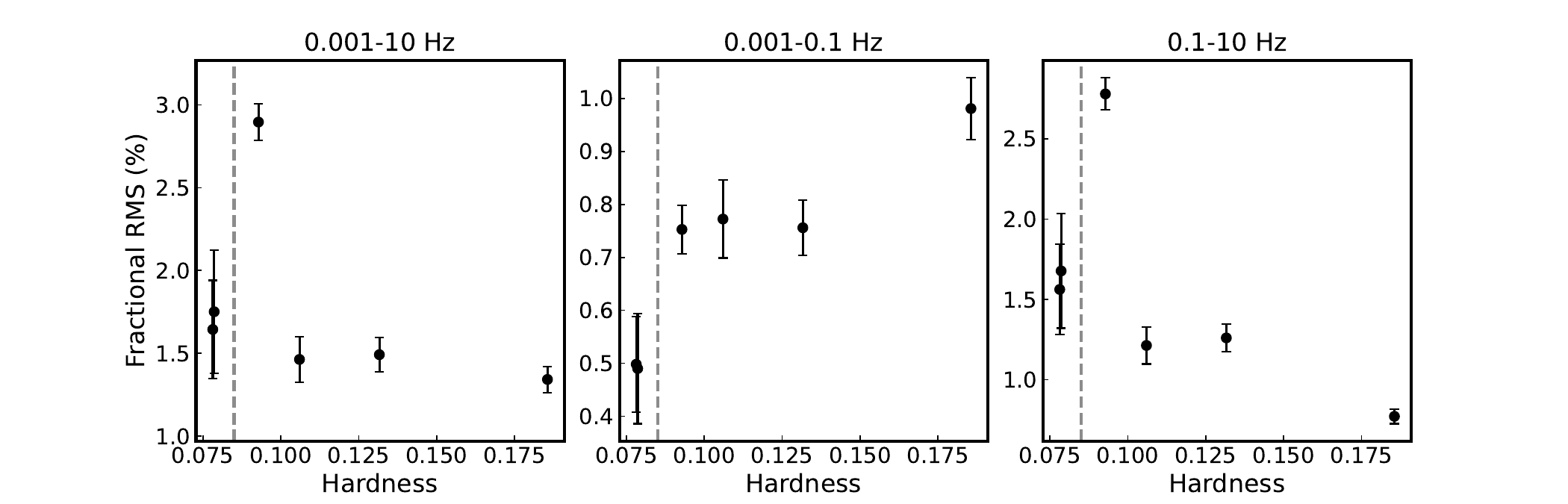}  
\caption{RMS vs. hardness evolution for the last six on-axis observations of 4U 1543--475 during its 2021 outburst. The left panel represents the integrated frequency range of 0.001-10 Hz, the middle panel represents the 0.001-0.1 Hz frequency range, and the right panel represents the relation for the 0.1-10 Hz frequency range. The vertical dashed grey line represents the state transition (VHS to SIMS) considered at the mid-point of the separation between obs5 and obs6.}\label{Fig:6} 
\end{figure*}

We have computed the integrated RMS fractional amplitude of variability for all nine observations. The RMS value was found to be low for all nine cases. However, there is a complex evolution observed across them. Background correction of RMS has been applied to neglect the effects of background contribution to the variability using the following formula.
\[r'=r \cdot \left( \frac{c}{c - b} \right) \]
where $r'$ is the background corrected RMS, $c$ is the total count rate, and $b$ is the background count rate. We have used the spectrum and background spectrum (see \S~\ref{sec:2}) to calculate the required count rates.
The left panel of Figure~\ref{Fig:5} represents the frequency-integrated (0.001--10 Hz) RMS evolution for all nine observations. The sudden jump in the integrated RMS for the eighth observation is the result of the BLN observed in the power spectra at higher frequencies. 

\subsubsection{\textbf{Frequency-resolved Variability:}} To further understand the origin of the variability components,  the evolution of frequency-resolved RMS amplitude was obtained. We have segmented the frequency range into 0.001-0.1 and 0.1-10 Hz. The middle and right panel of Figure~\ref{Fig:5} represents the RMS evolution for the 0.001-0.1 and  0.1-10 Hz range, respectively. After the fifth observation, we see a sudden jump in low-frequency RMS value, which is consistent with the sudden jump in hardness, which corresponds to a state transition from Very high state (VHS) to Soft intermediate state (SIMS). Table~\ref{Tab:2} represents the RMS values for different frequency ranges and the state of the source for all nine observations. As the count rate for the first three offset observations was underestimated due to the lower sensitivity (resulting from the effective area at the edge of the field-of-view), unusually higher values of low-frequency RMS, compared to that typically observed in the very soft state of BH XRBs, can be seen. Therefore, the focus has been directed towards the RMS evolution for the last six on-axis observations, enabling a comparison of the global timing and spectral evolution results. It is to be noted that the evolution of variability in the different frequency bands is distinctly different from each other. The low value of the high-frequency variability for the offset observations also indicates a small variability presence/detected compared to the on-axis observations.
\begin{table}
\begin{center}
\scalebox{0.7}{%
    \begin{tabular}{cccccc}
    \hline
   obs No. & &RMS (\%)& &  Spectral state  \\
   & 0.001-10 Hz&  0.001-0.1 Hz & 0.1-10 Hz&  of source\\
   \hline\\
   1 & 1.38 $\pm$ 0.16& 1.04 $\pm$ 0.12 &0.27$\pm$0.03 &VHS\\
   2&  1.06$\pm$0.12 & 0.80$\pm$0.09 & 0.21$\pm$ 0.02&VHS\\
   3 &1.78$\pm$0.25 & 1.16$\pm$0.16 & 0.12$\pm$0.02&VHS\\
   4 & 1.75$\pm$0.37 & 0.49$\pm$0.10 & 1.68$\pm$0.35&VHS\\
   5 & 1.64$\pm$0.29 & 0.49$\pm$0.09 & 1.56$\pm$0.28&VHS\\
   6 &1.46$\pm$0.14 & 0.77$\pm$0.07 & 1.21$\pm$0.12 &SIMS\\
   7& 1.49$\pm$0.10 & 0.76$\pm$0.05 & 1.26$\pm$0.09&SIMS\\
   8& 2.89$\pm$0.11 & 0.75$\pm$0.05 & 2.78$\pm$0.10&SIMS \\
   9 & 1.34$\pm$0.08 & 0.98$\pm$0.06 & 0.77$\pm$0.04&SIMS \\\hline
 
            \end{tabular}}
\caption{ Table of  fractional RMS amplitude for all nine observations in different frequency ranges
0.001-10 Hz, 0.001-0.1 Hz and 0.1-10 Hz. \label{Tab:2} }
\end{center}
\end{table}
\subsubsection{\textbf{Hardness vs RMS:}} 
To establish a relationship between hardness and RMS, hardness values have been computed using the energy ranges, following a similar approach as in the previous HID analysis (see \S~\ref{sec:3.1}). Utilizing the extracted spectra (see \S~\ref{sec:2}), we employed {\tt XSPEC} to calculate the count rate across different energy ranges, subsequently deriving the hardness values. It is important to note that we have utilized results exclusively from the on-axis observations for visualizing this relationship.
In Figure~\ref{Fig:6}, the left panel illustrates the RMS vs. hardness evolution across the entire frequency range, the middle panel showcases the relation within the 0.001-0.1 Hz frequency range, while the right panel displays the relation within the 0.1-10 Hz frequency range. The low-frequency RMS increases steadily with hardness, whereas the high-frequency RMS starts decreasing consistently with hardness after the state transition. The point of state transition is represented by the vertical dashed grey lines drawn at the mid-point in between obs5 and obs6 in Figure~\ref{Fig:6}. 
\subsubsection{\textbf{Energy-resolved RMS Evolution:}}
\begin{figure*}
\centering
\includegraphics[scale=0.45]{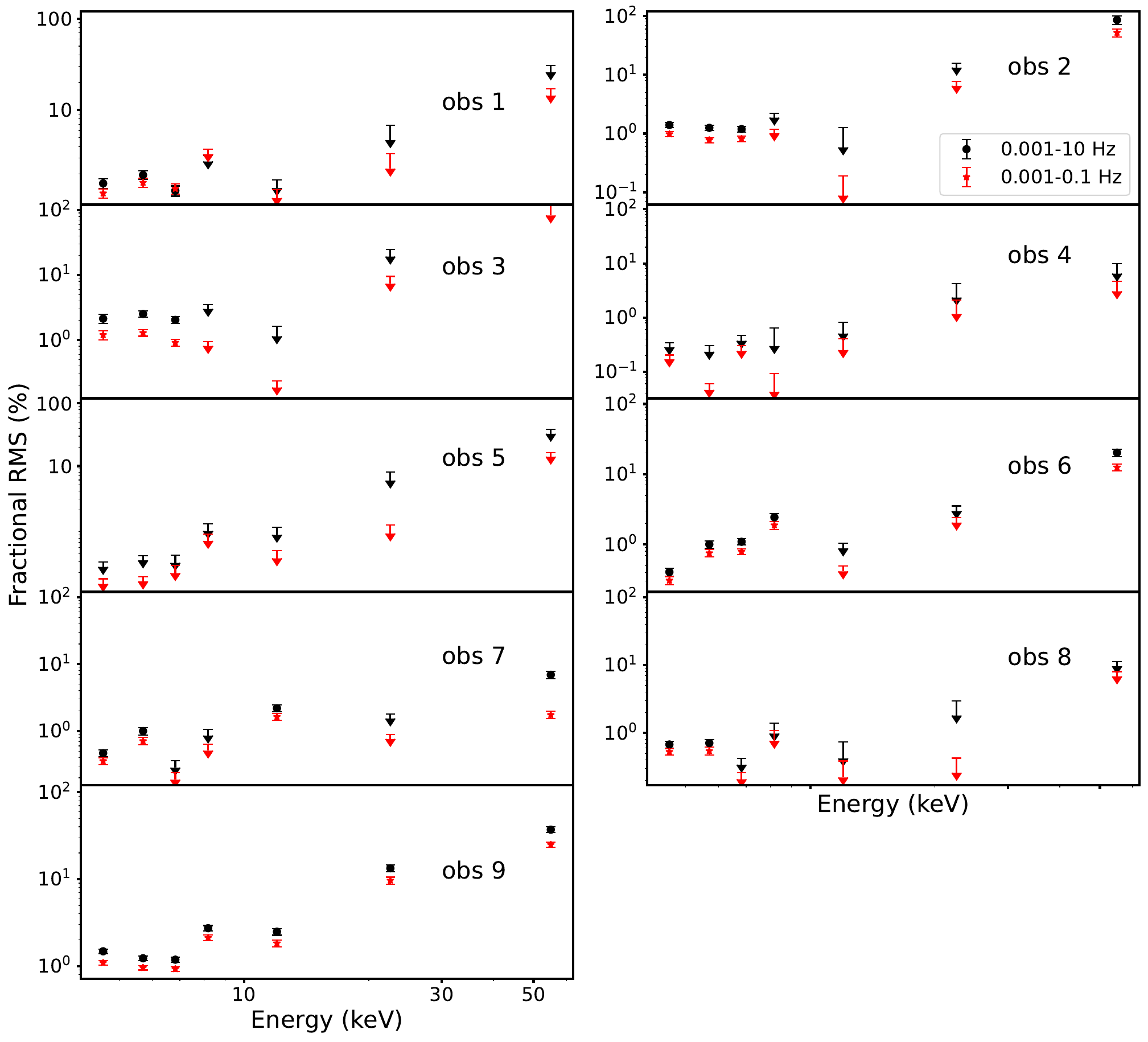}
    \caption{Energy and frequency-resolved RMS evolution for all the LAXPC observations of 4U 1543--475 during the 2021 outburst. Observations are represented by obs 1-9. The frequency-integrated RMS evolution is represented by black color, whereas red color represents the RMS evolution in 0.001-0.1 Hz. The tail of the arrows represents the upper limit of RMS. \label{Fig:7} }
\end{figure*}
For a deeper insight into the variability of the source, the investigation has been extended to fluctuations across the different energy bands: 4-5.14, 5.14-6.27, 6.27-7.38, 7.38-9, 9-15.11, 15.11-30 and 30-80 keV. Subsequently, Leahy normalized power spectra were generated for each energy range, employing logarithmic re-binning. 
Notably, an increased relative background contribution at higher energy levels has been observed due to the predominantly soft nature of the source during the outburst. Therefore, the intrinsic source variability was harder to detect at higher energies over the background. A red noise was visible for lower energy ranges, but was not clearly detectable for all high-energy power spectra. To further investigate these findings, the energy-resolved power spectra were fitted with two distinct models: one incorporating a constant model to represent Poisson noise and the other introducing a power-law component alongside the constant. 
And for a few cases, the power law contribution was weak. Therefore, we have used an F-test to validate the presence of the power law component. The power law contribution was considered for only those cases where the chance probability of the fit improvement on the addition of the power law component was less than 0.27\% (3$\sigma$ level). Upper limits (represented by the arrows) on the RMS values have been applied in the cases where this condition remains unsatisfied Figure~\ref{Fig:7}. The same approach has been applied across three frequency ranges: 0.001-10 Hz, 0.001-0.1 Hz, and 0.1-10 Hz.  illustrates the evolution of fractional RMS with energy across all nine observations. The black points represent the frequency-integrated RMS evolution, while the red points delineate the RMS evolution within the 0.001-0.1 Hz frequency range. In almost all instances, an observable uptrend in RMS with increasing energy emerges, signifying the noteworthy role played by higher-energy emissions in driving RMS variability. 

\subsubsection{\textbf{Time Lag Analysis}:}
Upon detecting substantial variations in RMS across multiple energy ranges, an investigation was performed to explore frequency-energy-dependent time lags across all nine observations following the method outlined by \cite{1999ApJ...510..874N}. The integration time of 1024 s and a Nyquist frequency of 128 Hz were adopted. The reference point chosen for this analysis was the lowest energy range, precisely 4-5.14 keV. This approach facilitated the computation of cross-spectra for the distinct energy intervals, 5.14-6.27, 6.27-7.38, 7.38-9 and 9-15.11 keV, from which time lags were computed. Intriguingly, the analysis unveiled time lags that were consistent with zero, indicating the absence of noticeable time delays across the considered energy ranges.
\begin{table*}[h!tb]
\begin{center}
\scalebox{0.77}{%
    \begin{tabular}{ccccccccccc } 
    \hline
     \hline
      Parameters & \multicolumn{9}{c}{Observation}\\
      & 1  & 2& 3 & 4 &5 & \textbf{6} & \textbf{7}&\textbf{8}&\textbf{9}  \\
      & offset ($40^\prime$) & offset ($40^\prime$) & offset($40^\prime$) & & && & &\\
    \hline
 T$_{in}$  (keV)& $1.076_{-0.005}^{+0.005}$ & $1.046_{-0.005}^{+0.005} $ & $0.990_{-0.005}^{+0.005}$   & $0.973_{-0.002}^{+0.002}$ &  $0.952_{-0.002}^{+0.002}$  & $0.948_{-0.002}^{+0.002}$ &$0.916_{-0.002}^{+0.002}$  &$0.903_{-0.001}^{+0.001}$&$0.873_{-0.002}^{+0.002}$   \\\\

          Norm$_{dbb}$   &  $5408_{-185}^{+190} $ & $4600_{-149}^{+151}$ & $5159_{-175}^{+215}$ & $  3921_{-83}^{+85}$  & $3985_{ -77}^{+90}$& $3744_{-79}^{+70}$ &$4525_{-81}^{+ 83}$ & $4463_{-75}^{+76}$ &$4739_{-88}^{+91}$\\ \\
 $\Gamma$ & $ 2.07 _{-0.09}^{+0.09} $ & $1.64_{l}^{+0.02}$ & $1.64_{-0.15}^{+0.12}$  & $1.77_{-0.08}^{+0.08}$ & $1.92_{-0.08}^{+0.08}$ &$1.85_{-0.03}^{+0.03} $&$1.85_{-0.03}^{+0.03}$ & $1.65_{-0.04}^{+0.04}$ & $1.88_{-0.03}^{+0.03}$ \\\\
        
  C$_{f}$  ($10^{-2}$ ) &$ 0.55_{-0.06}^{+0.07}$
 & $0.21_{-0.01}^{+0.01}$ & $0.20_{-0.03}^{+0.03}$ & $0.20_{-0.02}^{+0.05}$  & $0.22_{-0.02}^{+0.03}$ & $0.89_{-0.03}^{0.05}$ & $1.44_{- 0.05}^{+0.06}$ & $0.34_{-0.02}^{+0.02}$ & $2.47_{-0.09}^{+0.11}$\\\\
 E$_{Edge}$ (keV)  & $7.92_{-0.05}^{+0.05}$ &$7.89_{-0.04}^{+0.04}$&$7.77_{-0.07}^{+0.08}$ &$7.69_{-0.05}^{+0.06}$&$7.75_{-0.06}^{+0.04}$ &$7.45_{-0.08}^{+0.08}$&$7.29_{-0.10}^{+0.09}$&$7.51_{-0.06}^{+0.06}$&$7.14_{-0.19}^{+0.20}$ \\\\
 D &$0.85_{+0.04}^{+0.04}$ &$1.11_{-0.05}^{+0.05}$&$1.01_{-0.05}^{+0.05}$&$0.84_{-0.04}^{+0.04}$&$0.89_{-0.04}^{+0.04}$& ${0.79_{-0.04}^{+0.04}}$& $0.50_{-0.04}^{+0.04}$&$0.81_{-0.04}^{+0.04}$&$0.28_{-0.04}^{+0.04}$\\\\
 $Flux^{x}$  & $256.71_{-1.99}^{+2.03}$ &$180.23_{-1.68}^{+1.72}$&$140.95_{-1.15}^{+1.17}$&$94.95_{-0.74}^{+0.76}$ & $83.40_{-0.65}^{+0.66}$ &$78.76_{-0.62}^{+0.64}$&$77.25_{-0.61}^{+0.63}$&$65.23_{-0.51}^{+0.52}$&$60.55_{-0.49}^{+0.49}$\\\\
 Flux$_{thermal}^{x}$   & $252.28_{-1.98}^{+2.02}$ &$178.57_{-1.66}^{+1.69}$&$139.58_{-1.14}^{+1.17}$&$94.05_{-0.74}^{+0.76}$ & $82.57_{-0.65}^{+0.66}$&$5.52_{-0.61}^{+0.62}$&$71.88_{-0.59}^{+0.61}$&$63.92_{-0.50}^{+0.52}$&$53.17_{-0.46}^{+0.47}$\\\\
$Flux_{non-thermal}^x$   &$4.27_{-2.80}^{+2.87}$  &$1.78_{-1.91}^{+1.93}$ &$1.36_{-1.62}^{+1.66}$&$0.90_{-1.05}^{+1.07}$&$0.83_{-0.92}^{+0.94}$&$3.24_{-0.87}^{+0.89}$&$5.36_{-0.85}^{+0.88}$&$1.31_{-0.72}^{+0.74}$&$7.37_{-0.67}^{+0.69}$\\\\
\hline
Reduced $\chi^2$ (dof) & 1.21 (427) &  1.37(427) & 1.54 (427) & 1.52 (556) & 1.59 (556)  & 1.53 (556)& 1.67 (556) & 1.96  (556)&1.60 (556)\\
\hline
    \end{tabular}}
 \caption{Best fitted spectral parameters corresponding to the best-fitted model, {\tt tbabs*edge(diskbb*thcomp)} for all nine observations.
 \label{Tab:3}}
\begin{flushleft}
    \footnotesize{$T_{in}$: inner disk temperature}\\
    \footnotesize{$Norm_{dbb}$: disk blackbody normalisation}\\
    \footnotesize{$\Gamma$: power law index}\\
    \footnotesize{$^l$: lower limit for powerlaw index}\\
    \footnotesize{$C_f$: covering fraction}\\
    \footnotesize{$E_{Edge}$: line energy of the edge component}\\
    \footnotesize{$D$: absorption depth at the threshold}\\ 
    \footnotesize{$^x$  Flux level in units of $10^{-10}$ erg $s^{-1}$  $cm^{-2}$ in the energy range of 4 - 25 keV}
\end{flushleft}
\end{center}
\end{table*}
 
\subsection{\textbf{Spectral Analysis}\label{sec:3.3}}
We have conducted a joint spectral fitting using both LAXPC and SXT spectra for all nine observations to estimate the spectral evolution of the source during the decay phase of the outburst. For this analysis, \texttt{XSPEC} version 12.11.1 has been utilized, an integral component of the \emph{Heasoft} package version 6.29. Photons less than 4 keV and exceeding 25 keV for LAXPC, and photons less than 0.7 keV, and greater than 6 keV for SXT have been omitted from the analysis to maintain data integrity due to elevated background interference beyond this energy. Notably, the initial three observations were offset, necessitating the utilization of offset response files provided by the instrument team for LAXPC for precise spectral fitting during these observations. For the fit, 3$\%$ systematic errors have been used \citep{2022MNRAS.510.4040H, 2021MNRAS.505..713J}. To compensate for the effects of interstellar absorption, {\tt tbabs} model was used. The abundances and photoelectric cross-sections were adopted from \cite{2000ApJ...542..914W}.  The neutral hydrogen column density is used as 0.45  $\times$ $10^{22}$  cm$^{-2}$ \citep{2021ATel14725....1C}. Single component fits did not give us a good $\chi^2$  value, so we tried a two-component {\tt tbabs*(diskbb+powerlaw)} model initially, where {\tt diskbb} \citep{1984PASJ...36..741M} represents the thermal emission from the accretion disk which has two parameters, the inner disk temperature $T_{in}$ and the normalization which related to the inner disk radius.
The power law represents the non-thermal emission from the corona. The model gave us a good fit (reduced $\chi^2 <2$) with some noticeable residuals around 6-8 keV. We have added an {\tt edge} component with two parameters $E_{edge}$ (threshold line energy) and $D$ (absorption depth at the threshold energy) to take care of these residuals. The reduced $\chi^2$ value has been decreased further after adding the {\tt edge} model.
 \begin{figure}
    \begin{center}
\includegraphics[scale=0.53]{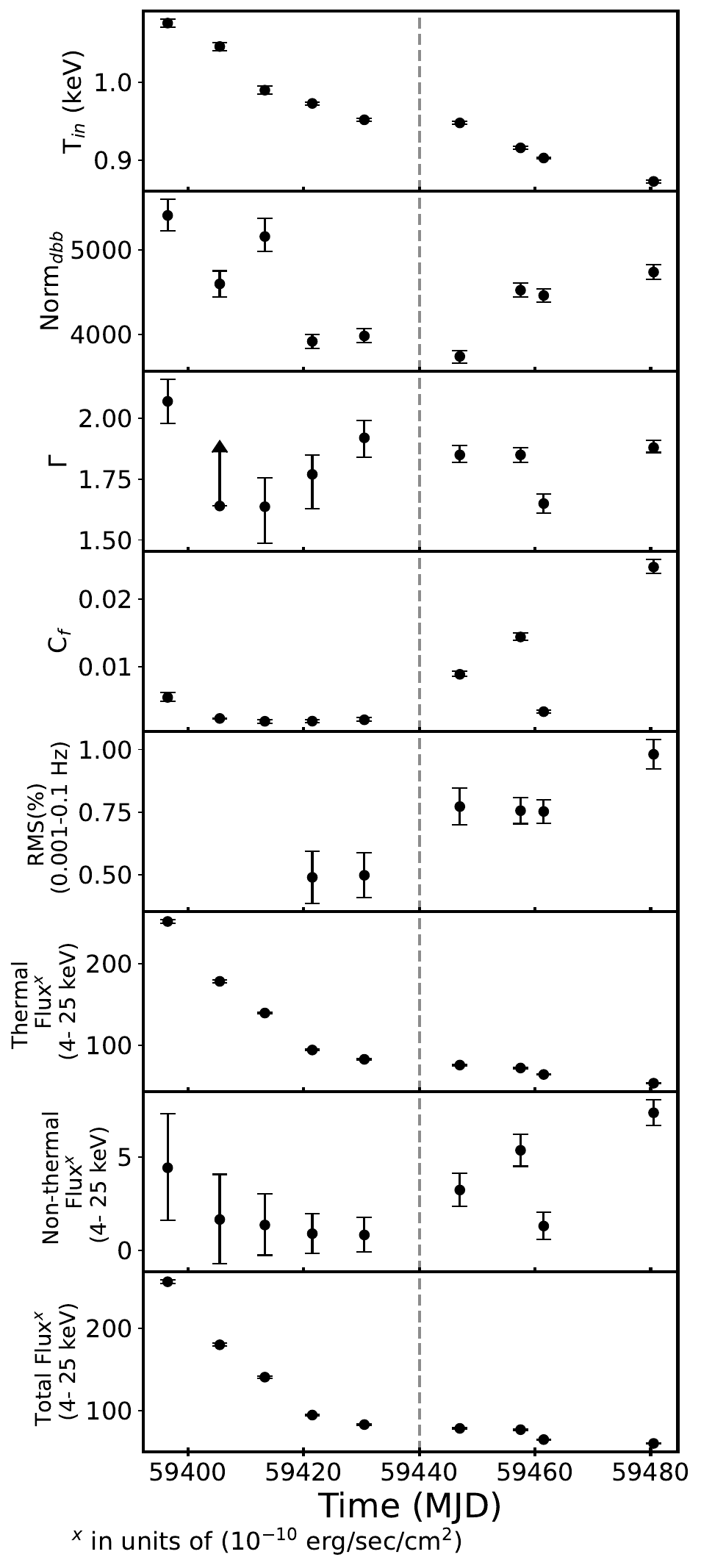}
    \caption{ Evolution of spectral parameters with time for all LAXPC+SXT observations of 4U 1543--475. First panel: inner disk temperature, second panel: disk black body normalisation, third panel: power law indices, fourth panel: covering fraction, fifth panel: low frequency (0.001-0.1 Hz) fractional RMS for the last six on-axis observations, sixth panel: thermal flux,   seventh panel: non-thermal flux,  eighth panel: total flux. The vertical dashed grey line represents the state transition (VHS to SIMS).
\label{Fig:8}}
\end{center}
\end{figure} 
We have included the cross-normalisation constant to address the uncertainties caused by cross-calibration of the SXT and LAXPC
instrument. We further explored several physical models, including {\tt compTT}, {\tt nthComp}, and {\tt thcomp}, to understand the non-thermal emission better. Among these, the {\tt thcomp} model emerged as the most suitable fit, leading us to adopt the { \tt tbabs*edge(thcomp*diskbb)} model.  The {\tt thcomp} model introduced by \cite{2020MNRAS.492.5234Z} characterizes a thermal comptonized continuum and involves four parameters: asymptotic power-law photon index ($\Gamma$), electron temperature ($kT_e$), covering fraction (a parameter influencing the extent of seed photon comptonization) and redshift. The electron temperature was kept fixed at 20 keV as mentioned by \citet{2023MNRAS.520.4889P}, as we could not constrain the parameter values considering $kT_e$ free. The results for the best-fitted model are reported in Table~\ref{Tab:3}. Figure~\ref{Fig:8} represents the evolution of the key spectral parameters during the decay of the outburst. We have also displayed the low-frequency RMS evolution for the last six on-axis observations in the fifth panel of the same figure to compare the spectral and timing evolution, as evolution is clearly visible in this particular frequency range.  A steady decrease of inner disk temperature has been observed from 1.076 $\pm$ 0.005 keV for the first observation to 0.873 $\pm$ 0.002 keV for the last observation, which is closely consistent with the result of \emph{NICER}- \emph{NuSTAR} joint spectral fit reported in \citet{2023MNRAS.520.4889P}. The normalization fluctuates during the off-axis observations, then it decreases to 3921 $\pm$ 85 for the fourth observation and remains almost constant till the sixth observation, followed by a sudden increase to 4525 $\pm$ 83.
 
A decreasing trend of the power law index has also been found for the first three offset observations (2.07$\pm$ 0.09 to 1.64 $\pm$ 0.13), then it increased again till the fifth observation (1.95 $\pm$ 0.08), and remained almost constant up till the seventh epoch. In the second offset observation, the power law index dropped unusually low, which appears nonphysical considering the source was in a softer state. Therefore, in this case, we applied a lower limit of 1.64.  We saw a sudden drop of the power law index for the eighth epoch (1.85 $\pm$0.03 to 1.65 $\pm$ 0.04), followed by an increase to 1.88 $\pm$ 0.03 for the last epoch.
A particularly noteworthy result is the evolution of the covering fraction. After an initial decay from  5.5 $\times $10$^{-3}$  to 2.1 $\times$ 10 $^{-3}$ , the covering fraction remains constant until the fifth observation. Subsequently, there was a sudden surge in the sixth and seventh observations to 8.9 $\times$ 10$^{-3}$ and 14.4 $\times$ 10$^{-3}$  respectively, which is an indication of state transition and is denoted by the dashed grey line is Figure ~\ref{Fig:8}. However, it decreased again in the eighth observation to 3.4 $\times$ 10$^{-3}$, only to rise once more to 24.7 $\times$ 10$^{-3}$  in the final observation. The evolution indicates that a higher number of seed photons are interacting with the corona in these epochs (sixth, seventh and ninth).
We have focussed on the spectro-temporal correlation of this source, and thus, to compare the correlated evolution of the variability (estimated using LAXPC data only) with the spectral changes in different stages during the decay phase of the outburst, we considered the thermal, non-thermal, and total flux values in the relevant energy range of LAXPC (4-25 keV).
The evolution of the thermal, non-thermal, and total flux in the energy range of 4-25 keV have been plotted in the last three panels of Figure~\ref{Fig:8}. The thermal flux in the energy range of 4-25 keV decreased steadily with time, but the non-thermal flux followed a similar trend as the covering fraction. 
\begin{figure*}
\centering\includegraphics[height=0.35\textheight]{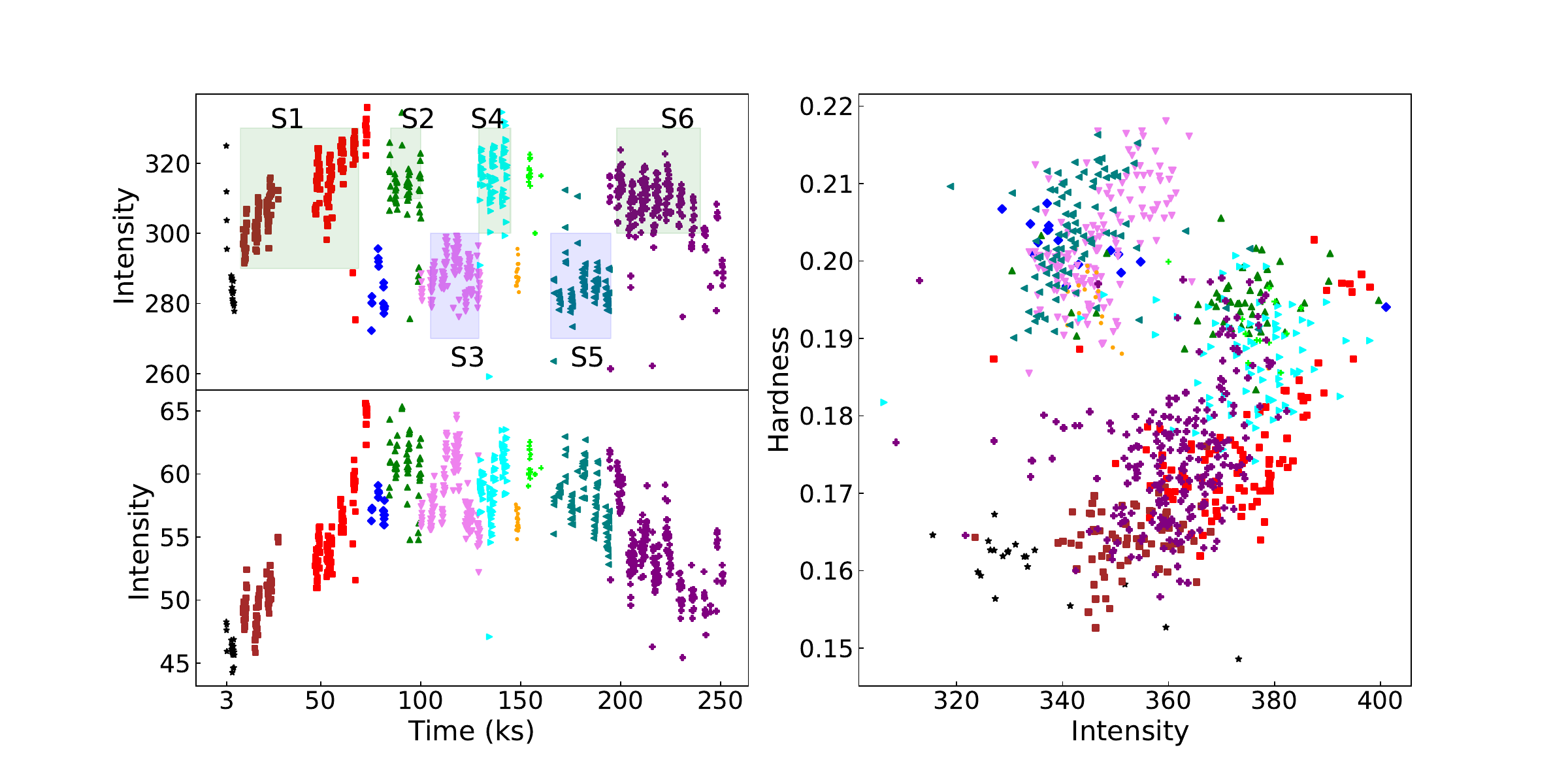}
\caption{The light curve and hardness intensity diagram during the ninth observation of \src  as observed by LAXPC. The upper left panel represents the light curve in the 5.14-8.46 keV energy range, and the lower left panel represents the light curve in the 8.46--18 keV energy range using a 100 s bin size. The right panel represents the hardness intensity (5.14--18 keV) diagram throughout the observation. Different colors represent different portions of the light curve to understand how the source evolves throughout the observation. The highlighted area represents the start and end times of the segments (S1, ..., S6) considered for spectral and timing analysis. Only the data in the highlighted regions was considered for the following spectral and timing analysis described in the subsequent subsections} \label{Fig:9}
\end{figure*}
\subsection{\textbf{Flip-Flop} \label{sec:3.4}}
In the ninth observation, we detected a recurring flip-flop pattern within the light curve, characterized by rapid oscillations in X-ray intensity. The upper left panel of Figure~\ref{Fig:9} illustrates the soft (5.14--8.46 keV) light curve, revealing a blend of steady evolution and abrupt fluctuation. 
Initially, over a span of approximately 75 ks, a consistent upsurge in soft  (5.14--8.46 keV) and hard (8.46--18 keV) photon counts have been noted. Subsequently, a sudden downturn in the count of soft photons ensued, leading to the source lingering in a lower flux level for around 8 ks. From there, the source transitioned to a higher flux level, maintaining this state for about 17 ks, only to revert to the lower flux level for a duration of 30 ks. Following this, a return to the higher flux level took place for approximately 17 ks, eventually leading to a remarkably rapid transition. The source experienced a swift shift from a higher to a lower flux level, promptly oscillating back to the higher flux level before settling again into the lower flux level for around 30 ks. Considering that the majority of these segments are separated by different orbital periods, precise estimation of the exact time intervals proves challenging. 
Nonetheless, we can discern a consistent order during which the source maintains a relatively steady flux level. 
The hardness intensity diagram has been plotted (right panel of Figure \ref{Fig:9}) following the same prescription mentioned in Section~\ref{sec:3.1}, the only difference is each data point corresponds to 100 s bin size. In the HID, we have observed two separate branches.
The light curve of the last observation has been divided into a few segments based on their position in the two branches in the HID to do the timing and spectral analysis. 
For a comprehensive depiction of the short timescale evolution of the source, we focused on the longer duration segments  ( $>$15 ks), which gave us reasonable power spectra to analyze the variability due to the fluctuations in the source.
Thus, six segments were depicted based on these criteria.
The flux level also varies whenever a jump from the lower to the upper branch has been noticed in the HID. Thus, the segments are separated accordingly. The segments are denoted by S1, S2, S3, S4, S5, and S6 in Figure~\ref{Fig:9} represented by different markers.  We considered different colors to represent the time segments exhibiting a change in the flux level in the light curve when the source remains in the new flux level for more than 1000 seconds after a transition. We have used two different colors for S1 just to see the evolution in HID during this evolutionary phase. After S5, a sustained decline in the count rate has been observed (S6). Sometimes, the transition in the light curve occurs in a single orbit file for those cases, we see a mixture in colors in different flux levels at the junction of two different segments. We have removed those contamination that occurred at the junctions during GTI selection, and thus, the highlighted boxes are smaller than the colored sections. 
We have not seen any clearly detectable flip-flop behaviour in the SXT light curve. corresponding to this observation. 
The non-detection of the significant flip-flop presence in the SXT light curve may be due to the poorer sensitivity of SXT compared to LAXPC, resulting from the significant difference in their effective areas. We have also checked the full energy band, 4-80 keV and 4-30 keV of LAXPC for the presence of the flip-flop. 
Both of these light curves show the presence of the flip-flop behaviour. However, to highlight the fact the flip-flop is primarily present only in the soft energy band (5.14-8.46 keV) and the flip-flop phase exhibits two distinct populations in the HID, we present just the light curves of 5.14-8.46 keV and 8.46-18 keV energy ranges in Figure~\ref{Fig:9}.
 
\begin{figure*}
\centering
    \includegraphics[scale=0.50]{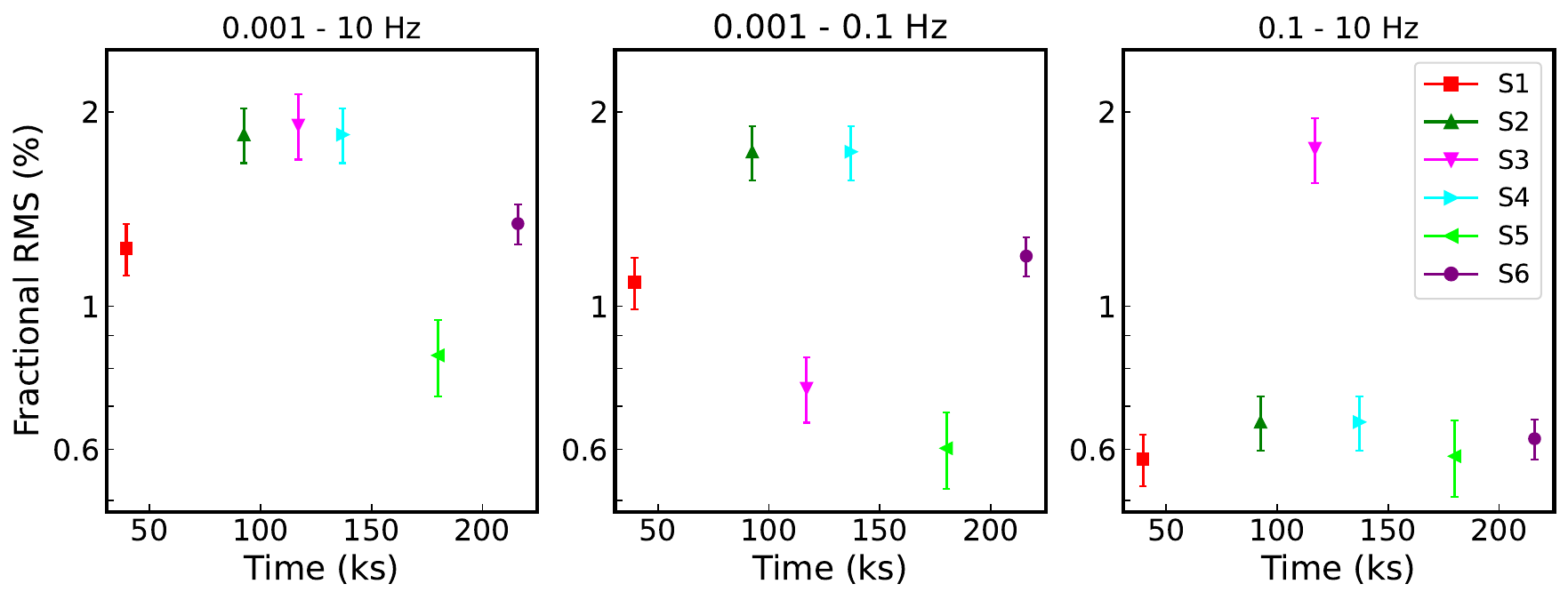}
\caption{Evolution of fractional RMS in different frequency ranges 0.001-10 Hz (left panel), 0.001-0.1 Hz (middle panel), 0.1-10 Hz (right panel) for the selected six segments of the light curve of the last observation. Different marked/colored data points represent different segments, as mentioned in the legend. \label{Fig:10}}
\end{figure*}
\subsubsection{\textbf{Temporal Analysis of Flip-Flop}:}
A power spectral analysis has been conducted to explore the temporal characteristics of the flip-flop event. The power spectra were generated for the six selected segments using methodologies similar to those outlined in Section  \S~\ref{sec:3.2}. We have used the entire 4-80 keV energy range for the variability analysis of the flip-flop phase. In the resulting power spectra, a power law noise in the lower frequency range, along with the Poisson noise, was observed. The power law model (M1) was fitted to the Poisson noise-subtracted power spectra. The RMS variability was then calculated using the procedure detailed in Section \S~\ref{subsec:3.2.1}.
The general nature of the intrinsic source variability has been found to be the same between the two different flux levels; however, the magnitude of variability showed a variation.
A frequency-resolved RMS analysis has been performed to discern the origin of this variability, calculating the RMS for two distinct frequency ranges: 0.001-0.1 Hz and 0.1-10 Hz. The progression of RMS values across different frequency ranges is illustrated in Figure~\ref{Fig:10}. Notably, the alteration in RMS displays variability that differs among distinct frequency ranges. Given that the power spectra are primarily dominated by Poisson noise power at higher frequencies, our focus for investigating source variability is directed towards the 0.001-0.1 Hz frequency range. The analyzed segments have been categorized into three groups: the evolutionary stage (S1 \& S6), higher flux level (S2 \& S4) during flip-flop and lower flux level (S3 \& S5) of flip-flop. During the evolutionary stage, the RMS variability (0.001-0.1 Hz) remains nearly constant, taking into account the error bars. The RMS values are 1.09 $\pm$ 0.10 $\%$ and 1.19 $\pm$ 0.09 $\%$, respectively. Conversely, the four segments situated in the middle showcase a distinct flip-flop phenomenon.
The RMS values undergo considerable alteration during the transition from a higher flux level to a lower flux level (S2 to S3 and S4 to S5). 
In the case of S3, a substantial reduction in the count rate accompanied by a shift in hardness towards the higher branch has been observed. Coinciding with this behavior, the power law index of the power spectra also exhibited a decrease (1.42 $\pm$ 0.07 to 0.64 $\pm$ 0.05). This change increased the area under the curve at higher frequencies, leading to heightened variability. This phenomenon elucidates the sudden surge in the RMS value at the higher frequency range (0.1--10 Hz) for the third segment, which is evident in the right panel of Figure~\ref{Fig:10}. 
We have also investigated the variability in an intermediate frequency range (0.1-1 Hz), and there is no correlated behaviour of the RMS in this frequency range is observed with the flux transitions, possibly because the source variability gets buried under the Poisson variability even for this frequency range in these soft states of the source.
\begin{figure}
\centering
 \includegraphics[scale=0.38]{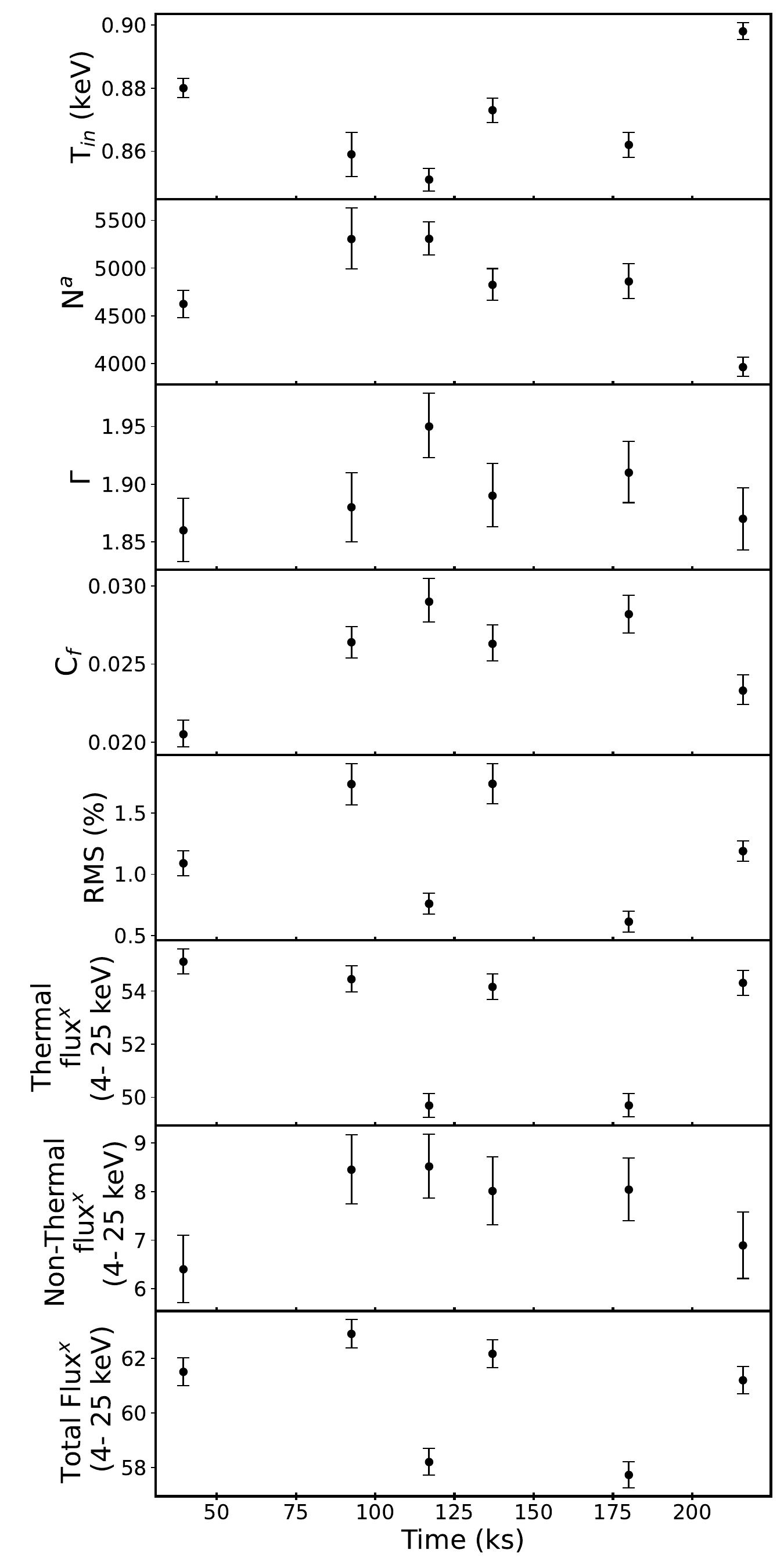}
\caption{Evolution of spectral parameters for the six segments of the last observation of 4U 1543--475. First panel: inner disk temperature,  second panel: disk black body normalization, third panel: power-law indices, fourth panel: covering fraction, fifth panel: Fractional RMS (0.001 - 0.1 Hz), sixth panel: thermal flux, seventh panel: non-thermal flux, eighth panel: total flux. \label{Fig:11}}
\end{figure} 
\subsubsection{\textbf{Spectral Analysis of Flip-Flop}:}
The spectra were extracted and modeled for each of the six distinct selected segments of the ninth observation. A similar method has been used to those mentioned in \S~\ref{sec:3.3}. In our joint spectral analysis, each spectrum was meticulously fitted utilizing a combination of thermal and non-thermal models, specifically {\tt tbabs*edge(thcomp*diskbb)} as employed previously. Figure~\ref{Fig:11} represents the evolution of the joint spectral parameters as well as the low-frequency RMS along with the evolution of thermal, non-thermal, and total flux in the 4-25 keV energy range during the decay of the outburst. We have observed the flip-flop phenomenon only in the LAXPC data; SXT light curve does not show flip-flop behavior, most probably due to less sensitivity; thus, to examine the spectro-temporal evolution, we have focused on the flux of 4-25 keV energy range (LAXPC spectral sensitivity range) for the flip-flop analysis.
Our flip-flop analysis yields compelling insights into the evolution of the source. During the transition from S1 to S2, we observe an increase in disk normalization, indicating a shift of the inner disk radius away from the compact object and, consequently, a decrease in the inner disk temperature. The power law index remains consistent, $1.86 \pm 0.03$ for S1 and $1.88 \pm 0.03$ for S2. While the thermal flux for 4-25 keV energy range remains almost constant, the covering fraction sees a notable rise from ($2.05 \pm 0.08$)$\times$ 10$^{-2}$  to ($2.64 \pm 0.10$ )$\times$ 10$^{-2}$.
Examining the HID diagram, we observe a gradual growth in hardness from S1 to S2, indicative of a global evolution of the source. From S2 to S3, the disk normalization remains constant, with a subtle hint of a decrease in the inner disk temperature, although not significant. The non-thermal flux in the 4-25 keV energy range remains constant while a sudden drop in thermal flux is noticed, which supports the reduction in the inner disk temperature. The power law index experiences a slight increase from 1.88 $\pm$ 0.03 to 1.95 $\pm$ 0.03, accompanied by a rise in the covering fraction from ($2.64 \pm 0.10$)$\times$ 10$^{-2}$ to ($2.90 \pm 0.15$) $\times$10$^{-2}$.  
A pronounced transition is observed from S3 to S4, marked by a decrease in the inner disk radius, leading to an elevation in the inner disk temperature to $0.873 \pm 0.004$ keV. Consequently, the thermal flux in the 4-25 keV energy range increases, returning to the levels observed in the S1 and S2 segments. While there is a moderate decrease in covering fraction and power law index, the non-thermal flux remains unchanged (Figure~\ref{Fig:11}).
Moving from S4 to S5, the inner disk radius remains constant, but a decrease in temperature and a resulting reduction in the thermal flux (4-25 keV) is noticed. Notably, there is no substantial change in covering fraction and power law index, although the RMS variability decreases.
In the transition from S5 to S6, the disk approaches the compact object closely, causing the temperature to reach $0.898 \pm 0.003$ keV. The power law index remains constant, but there is a significant decrease in covering fraction. The thermal flux (4-25 keV) increases to levels observed in S1, S2, and S4. The non-thermal flux (4-25 keV) remains nearly constant, while the RMS variability increases once again.
In the SXT light curve and flux evolution, there is an absence of a flip-flop pattern. However, there is a noticeable evolution in the SXT flux (0.7-5 keV) level between adjacent segments. Specifically, the thermal flux (0.7-5 keV) decreased from ($482.45 \pm 2.86$) $\times$ 10$^{-10}$ to ($454.36 \pm 1.47$) $\times$ 10$^{-10}$ erg s$^{-1}$ cm$^{-2}$ as the source transitioned from S2 to S3. 
Subsequently, there was a modest increase in thermal flux to ($457.91 \pm 1.55$) $\times$ 10$^{-10}$ erg s$^{-1}$ cm$^{-2}$ as the source entered S4. 
However, during the S4 to S5 transition, there was a drop in the thermal flux level, reaching ($437.41 \pm 1.61$) $\times$ 10$^{-10}$ erg s$^{-1}$ cm$^{-2}$. 
Interestingly, the non-thermal flux in SXT (0.7-5 keV) remained consistent with that in LAXPC (4-25 keV) during the flip-flop.

\begin{table}[h!tb]
\begin{center}
\scalebox{0.75}{%
    \begin{tabular}{cccc } 
    \hline
     \hline
      Parameters & & Q1 & Q2   \\
      \\
    \hline
 T$_{in}$  (keV)&&0.869$_{-0.004}^{+0.004}$ & 0.855$_{-0.003}^{+0.003}$  \\\\
 Norm$_{dbb}$&&4948$_{-157}^{+163}$  &5122$_{-142}^{+146}$\\ \\
 $\Gamma$ &&1.89$_{-0.03}^{+0.03}$ & 1.93$_{-0.03}^{+0.03}$ \\\\
C$_{f}$  ($10^{-2}$ )&&2.64$_{-0.11}^{+0.12}$ &2.91$_{-0.12}^{+0.14}$ \\\\
 E$_{Edge}$ (keV)& &7.28$_{-0.23}^{+0.23}$ &7.14$_{-0.26}^{+0.27}$\\\\
 D &&0.23$_{-0.04}^{+0.04}$&0.20$_{-0.04}^{+0.04}$\\\\
 $Flux^{x}$ &(4-25 keV)&62.36 $_{-0.50}^{+0.51}$ &57.93$_{-0.47}^{+0.48} $ \\\\
 &(0.7-5 keV)&468.19$_{-1.48}^{+1.48}$ &452.49$_{-1.25}^{+1.25}$ \\\\
 Flux$_{thermal}^{x}$ &(4-25 keV)& 54.22$_{-0.47}^{+0.48}$ &49.64$_{-0.44}^{+0.45} $ \\\\
  &(0.7-5 keV)&463.21$_{-1.46}^{+1.46}$ &447.23$_{-1.23}^{+1.25}$ \\\\
$Flux_{non-thermal}^x$&(4-25 keV) &8.14$_{-0.69}^{+0.71}$ & 8.29$_{-0.64}^{+0.66}$ \\\\
  &(0.7-5 keV)&4.97$_{-2.07}^{+2.08}$ &5.26$_{-1.75}^{+1.76}$ \\\\
Reduced $\chi^2$ (dof)&&1.25 (456)&1.08 (456) \\\\
\hline

RMS (\%)&0.001-10 Hz&1.49 $\pm$ 0.14&1.31$\pm$0.15\\
&0.001-0.1 Hz&1.17 $\pm$0.11&0.69 $\pm$ 0.08\\
&0.1-10 Hz &0.57 $\pm$ 0.05 &1.09 $\pm$ 0.12\\
\hline
    \end{tabular}}
 \caption{Best fit spectral parameters corresponding to the model, {\tt tbabs*edge(diskbb*thcomp)} and the RMS values for three different frequency ranges for the two sets of combined segments, S2 with S4 (Q1, merged upper flux level), S3 with S5 (Q2, merged lower flux level). 
 \label{Tab:4}}
\begin{flushleft}
    \footnotesize{$T_{in}$: inner disk temperature}\\
    \footnotesize{$Norm_{dbb}$: disk blackbody normalisation}\\
    \footnotesize{$\Gamma$: power law index}\\
    
    \footnotesize{$C_f$: covering fraction}\\
    \footnotesize{$E_{Edge}$: line energy of the edge component}\\
    \footnotesize{$D$: absorption depth at the threshold}\\ 
    \footnotesize{$^x$  Flux in units of $10^{-10}$ erg $s^{-1}$  $cm^{-2}$ in two different energy ranges }
\end{flushleft}
\end{center}
\end{table}

\subsubsection{\textbf{Analysis of Merged Segments} }
In the time-resolved analysis of the flip-flop segments, we could not identify significant systematic variations in the spectral parameters.  It is to be noted that the flip-flop phenomenon could not be clearly detected in the SXT light curve, probably due to lower sensitivity (effective area) of the SXT compared to LAXPC.
Therefore, to enhance significance, we grouped the nearby segments of the light curve which lie in the same branch in the HID. We grouped S2 with S4 (Q1) and S3 with S5 (Q2), assuming they exhibit similar emission behavior. This assumption is supported by the fact these segments at each flux level show similar behavior with regard to their intensity and color. This helps us delve into the physical processes causing the source to transition between two branches in the HID. We then extracted power spectra from LAXPC data for these combined segments. Using the method described in \S~\ref{sec:3.2}, we conducted RMS analysis to understand the timing properties of the source in the two branches of HID. Furthermore, we extracted the spectra for LAXPC and SXT for each of these combined segments and performed joint spectral analysis following the model and method mentioned in \S~\ref{sec:3.3}. For both combined segments, we considered single-pixel events to extract SXT spectra.
Table \ref{Tab:4} represents details of the timing and spectral parameters for two different combined segments. We have observed the temperature decreased significantly  \textbf{($>$3 $\sigma$) } between these two branches from 0.869 ± 0.003 keV to 0.855 ± 0.003 keV while the inner disk radius and the power law index remain constant considering the error bars. We have observed a slight increase in covering fraction from ($2.64 \pm 0.12$)$\times$ 10$^{-2}$  to ($2.91 \pm 0.14$) $\times$ 10$^{-2}$. The thermal flux and total flux for both energy ranges (0.7-5 keV and 4-25 keV) changed significantly, whereas non-thermal flux remains constant. We have also seen a significant decrease in the low-frequency RMS from 1.17 ± 0.11 \% to 0.69 ± 0.08 \% and a significant increase in high-frequency RMS from 0.57 ± 0.05 \% to 1.09 ± 0.12 \%. 

\begin{figure}
\centering
\includegraphics[scale=0.55]{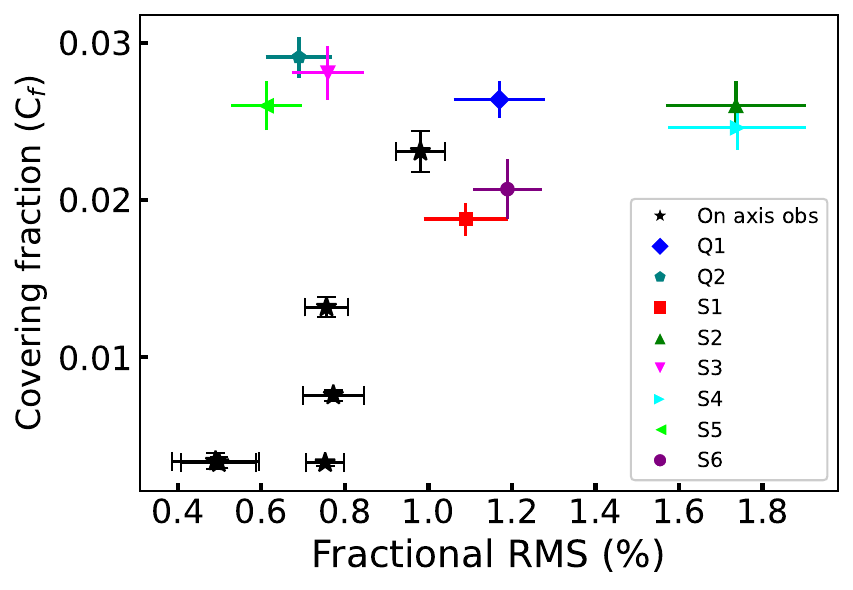}
\caption{ Evolution of covering fraction ($C_{f}$) with fractional RMS (0.001 - 0.1 Hz) of 4U 1543-475 for 2021 outburst. The black data points represent the on-axis observations, whereas the different colored/marked data points represent the different segments of the last light curve.\label{Fig:12} }
\end{figure}

\section{DISCUSSION \& CONCLUSION \label{sec:4}}

In this work, we conducted a spectro-temporal study of the BH XRB \src \ during its 2021 outburst. It is the brightest outburst ever observed from the source, reaching a flux level of $\sim$ 9 Crab in the 2-10 keV energy range \citep{2023arXiv230708973Y}.  From the temporal analysis, we have observed that except for the eighth observation, all the power spectra are consistent with a simple power law variability, which indicates the source is mostly in a high soft state during these epochs. An excess power at higher frequencies arises during the eighth epoch, modeled with a zero-centered Lorentzian (Figure~\ref{Fig:3}). The timescales corresponding to these low-frequency variabilities can be related to the inward propagation of the disk fluctuations. Typically, in the soft spectral state, the power density spectra (PDS) mostly exhibit a PLN \citep{1994ApJ...435..398M, 1997ApJ...487L..73C, 2006HEAD....9.0186S}. A BLN component has also occasionally been observed in addition to the PLN \citep{2021RAA....21..170M}. \cite{2013ApJ...770..135Y} associated the PLN component with the thermal disk component, which implies PLN is a signature of the variabilities originated in the standard accretion disk. The low variability and the power law shape of the power spectra in the soft state can be explained by the viscosity fluctuation in the accretion disk \citep{1997MNRAS.292..679L}.
In this work, whenever an increase in hardness has been encountered (Figure~\ref{Fig:2}), a corresponding evolution has been noticed in the low frequency (0.001 - 0.1 Hz) RMS evolution as well (see middle panel of Figure~\ref{Fig:5}). For example, the sudden increase of RMS for the sixth observation in 0.001-0.1 Hz is concurrent with an increase in the non-thermal emission (Figure~\ref{Fig:8}) and a drop in the disk fraction, which corresponds to a state transition. It is to be noted that the RMS for the first three offset observations appears to be higher (1.14 $\pm$ 0.14 \% for the first observation) than the RMS values during the harder state due to the underestimation of the intensity resulting from the offset.
To confirm this, we obtained the best-fit spectral model using the offset responses and convolved it with the on-axis responses to estimate the true count rate.
A significant increase is detected between observed (offset) and estimated true (on-axis) count rates (953.41 to 4352.91 counts/s, respectively, for the first observation). Using the estimated true count rate, the modified RMS values ($0.43\pm 0.05$ \% for the first observation) seem to fall in the same ballpark as the low RMS values observed for the on-axis observations in the high soft state.  A similar result is found by \cite{2005ApJ...622..508K}, where during the first state transition of the 2002 outburst from the thermally dominant (TD) state to the Intermediate state (IS), a rise in the RMS amplitude of variability and power-law flux is observed. They explained the scenario using the Advection dominated accretion flow (ADAF) model and the `magnetic-corona' model. In the latter case, the position of the inner edge of the disk does not vary significantly and the variation of the optically thin corona above the disk drives the spectral evolution of the source.

In our study, we have observed an increase in disk fraction 93.05 \% to 97.99 \% in 4-25 keV energy range as the source moves from obs7 to obs8.
This behavior is further consistent with the evolution of the disk reported by \cite{2023MNRAS.520.4889P}.
They studied the equivalent width of the new absorption feature observed in the \emph{NuSTAR} spectra (see Figure 9 of \cite{2023MNRAS.520.4889P}) and inferred a sudden infall of disk materials (around an epoch corresponding to our sixth observation) leading to an evacuation of the inner disk and then a disk recovery phase for the next 10 days. Such a scenario involving disk recovery following the infall phase can well explain the enhanced disk fraction during the eighth epoch.
These findings suggest that the ``magnetic-corona" model may not adequately explain this situation and argue for an alternative scenario. After the state transition in the sixth observation, the appearance of BLN is observed in the power spectra of the eighth observation during the disk recovery phase. This is consistent with the general trend of the changes in the variability spectrum with state transition in the case of low-mass X-ray binaries \citep{2022ApJ...930...18W}. As the source transitions from a soft to a hard state, the emergence of intricate high-frequency fluctuations is observed in the power spectra. 
Finally, an increase in low-frequency RMS in the ninth observation is noticed, which is consistent with the corresponding increase in hardness.

A combined energy and frequency-resolved RMS analysis is performed to understand the origin of the variability. However, we could not validate the presence of the PLN or the intrinsic source variability components for all finer energy bands except for the ninth epoch after using the F-test due to a lack of source variability at higher energy. The very low-frequency variability is weakly detectable for the higher energies, particularly in the softer states. This may imply that the very low-frequency variability in the high soft state is governed predominantly by the component responsible for the thermal emission, i.e., the accretion disk.
The RMS shows an increasing trend with energy for almost all cases, considering the upper limits and error bars. Specifically, for the harder observations (obs 6 and obs 9), the variability could be identified with 3$\sigma$ significance over a broadband range, suggesting an increasing contribution of the higher energy component (non-thermal) to the variability for these epochs. Such an increasing trend of RMS with energy is consistent, as observed in other BH XRBs. Similar result is found by \cite{2005MNRAS.363.1349G} for XTE J1550–564 and XTE J1650–50, \cite{2020MNRAS.497.3896A} for MAXI J1727–203, and \cite{2021MNRAS.505..713J} for MAXI J1348–630. Some neutron star XRBs show similar behavior (e.g. XTE J1701-462; \citep{2015ApJ...799....2B}, GX 349+2; \citep{2023MNRAS.523.2788K}).
This scenario can be explained by the radiative process in the corona, which can amplify the variability originating at the disk, leading to an enhanced RMS strength with energy, especially for the cases where the corona is significantly contributing \citep{2021ASSL..461..263M}.
A direct correlation between the timing and spectral properties of the XRB can be observed from our results. To further probe the spectro-temporal correlation,  we have plotted the low-frequency RMS variability with covering fraction for the last six on-axis observations in Figure~\ref{Fig:12}. The covering fraction displayed a monotonically increasing behavior with the low-frequency RMS, especially following the spectral state transition to the intermediate state.

As the thermal flux decreases, it has been observed that the decreasing number of seed photons is inadequate to cool the corona, and a possible increase in coronal temperature might enhance the interaction probability and, consequently, lead to a rise in the covering fraction, which is evident from our result during the state transition as well as in the ninth observation (Figure~\ref{Fig:8}). During the eighth observation, we observed the opposite behavior. When the disk fraction increased during the recovery phase of the sudden infall of the inner disk, the covering fraction decreased. However, we could not constrain the electron temperature in our study. The increase in interaction might also be a result of the variation in the geometry of the corona.
Increasing the volume of the corona will again result in an increase in the covering fraction. 
The latter scenario can be related to the variable corona model \citep{2019Natur.565..198K}. If the corona expands vertically due to radiation pressure, the surface area will increase as well as the solid angle between the inner region of the disk and the corona. Thus, more photons can now interact with the corona and get comptonized, resulting in an increase in covering fraction and RMS variability as well.  
However, we are unable to infer anything about the length scale of the corona, as we could not detect any time lag between soft and hard photons for any observation that corresponds to the geometric delay. In the soft state, the origin site of the seed photons and the comptonizing region are substantially close to the compact object, resulting in a very low lag, which is difficult to detect, given the instrumental uncertainties.

In the MAXI light curve (Figure~\ref{Fig:1}), it can be observed that after the outburst peak, the flux continuously decays, followed by a dip around 31st August (MJD 49455) and then exhibits a sudden increase in flux. After attaining a small peak around 20th September, the flux again exhibits a gradual decline.
One of the LAXPC observations is synchronous with the occurrence of the sudden flux increase phase.
The light curve of this observation exhibits a phenomenon called flip-flop, which remains an enigma to date. Though the soft LAXPC light curve showed the flip-flop behavior, a clear signature of the flip-flop phenomenon could not be detected in the SXT light curve, possibly due to its poorer sensitivity (effective area).
To further understand how the source is evolving during the flip-flop, we conducted a time-resolved timing and spectral study of the segments of the light curve. 
The steady increase in count rate during the initial stage of the observation indicates an initial evolution of the source, which eventually ends up triggering the system and gives rise to the flip-flop behavior lasting till 200 ks (Figure~\ref{Fig:9}). The non-thermal flux (4-25 keV) remains almost constant during the flip-flop (Figure~\ref{Fig:11}). But there was a significant decrease in thermal flux (4-25 keV) when the source shifted from the higher to lower flux level, signifying an intriguing connection between flux variations and the thermal component. As a result, we have seen a spectral hardening as the flux level goes down. Similar hardening has been observed by \cite{1998ApJ...494..753K,2005ApJ...623..383H} while the sources were showing flip-flop like temporal characteristics in their light curves. \cite{2001ApJS..132..377H} observed a decrease in hard color in the very high soft (VHS) state while the flux level increased during the sudden flux transition phase. Though the soft color remained almost constant, there was a clear indication of a change in the inner disk radius by a few percent. From their results, they argued that the flux transition could be governed by the accretion disk component.  
In our result, the disk temperature displayed an indication of a decreasing behavior for the first transition and a moderate change later on (Figure~\ref{Fig:11}). However, this trend could not be confirmed to be statistically significant as there is an indication of the global evolution of the source along with the flip-flop. We have seen the thermal flux also shows significant change during the flux transition, which indicates and supports the evolution of the inner disk temperature. We have found a significant change in the inner disk temperature between the merged segments of flip-flop (Table ~\ref{Tab:4}). This suggests that the flip-flop behavior is regulated by variations in the accretion disk, aligning with the aforementioned findings.  
\citet{1994ApJ...435..398M} found that these rapid time variations of flux during flip-flop are due
to the comptonized component. \cite{2003A&A...412..235N} observed a 9\% decrease in thermal flux and a 30\% increase in non-thermal flux, resulting in spectral hardening as the source went into a higher flux level.    
This scenario is contrary to our results (Table ~\ref{Tab:4}), where the flip-flop phenomenon is primarily governed by the thermal component as non-thermal flux remains constant during the phase transition. 
It is evident from Figure \ref{Fig:10} that for the lower flux level flip-flop segments, the low-frequency variability has decreased, which is concurrent with a decrease in the disk fraction and a largely unvarying non-thermal flux. We have obtained similar results from our merged segments analysis. This indicates that the short timescale changes in the low-frequency variability are governed by the disk evolution, indicating the disk as the primary origin of the low-frequency power-law noise variability. 
\cite{2020A&A...641A.101B} noticed that a variation in the inner disk temperature of the accretion disk is responsible for most of the flux transitions during the flip-flop. \cite{2021MNRAS.505..713J}  also found flip-flop-type behavior in MAXI J1348–630 and observed variation in the spectral parameters consistent with our result (Table ~\ref{Tab:4}). In particular, the scattering fraction and the inner disk temperature exhibited changes during the flip-flop event.
\cite{2004ApJ...610..378P} reported hard dips during the steep power law (SPL) state of the 2002 outburst of \src. The dips lasted for 5-10 s, and the dip occurrence rate was roughly one per minute. They also suggested accretion instability for the occurrence of the dips in those states. The radiation pressure instability can act as a notable factor contributing to the short-timescale variability, as highlighted by \citet{2007A&ARv..15....1D}. The interplay between advection and radiation pressure can also trigger the flip-flop behavior.
Notably, GRS 1915+105 exhibits behavior (class $\lambda$, class $\kappa$) that closely resembles limit cycles in certain light curves, as noted by \citet{2000A&A...355..271B}.

We also investigated the validity of the relationship between the RMS and the covering fraction over the individual flip-flop segments as well as the merged segments (Q1 and Q2) to probe the contribution of the non-thermal component to the variability over these short timescales. The corresponding relationship is shown by the colored data points in Figure~\ref{Fig:12} for the different segments of the flip-flop phase of \src.
The segments corresponding to the upper flux levels in the flip-flop observation manifest an evolution consistent with the long-term evolution (represented by the black points) over the different observational epochs. Particularly, an increasing trend between the low-frequency variability and covering fraction is maintained following the onset of the intermediate state and through the high flux level intervals of the flip-flop observation.
This correlation underscores that an increase in photon-corona interaction corresponds to heightened variability, implying that the corona largely governs the evolution of the variability, especially after the high-soft state.
A distinctly different behavior is observed over shorter timescales during the flip-flop phenomenon. In the intermediate state, with increasing non-thermal parameters like covering fraction and flux (and consequently hardness), the variability shows a steady, systematic increasing trend. However, in the flip-flop phase, when the flux flips to the lower level, the variability exhibits a significant sudden disparity over these short timescales, while the non-thermal flux and components remain approximately non-variant, as displayed by the blue (Q1) and teal (Q2) point in Figure~\ref{Fig:12}. Such a scenario can possibly be attributed to the triggering of an instability in the thermal accretion disk that dampens the physical mechanism responsible for the low-frequency variabilities originating in the disk. 
In principle, such a flip-flop characteristic can be envisioned as a damped oscillator in the disk with an irregular interval \citep{2020A&A...641A.101B}. A mechanism is required to trigger the imbalance in the disk to create such oscillations between two unstable states. The energetic winds, with their elevated velocities, could potentially interact with the accretion flow \citep{2011IAUS..275..290N}, thereby altering the viscosity of the material accreting onto the inner disk. This alteration in viscosity can act as a potential triggering mechanism that underpins the generation of the flip-flop phenomenon.

\section{ACKNOWLEDGEMENTS}
This study incorporates data collected through the \emph{AstroSat} mission. We acknowledge the ISSDC (Indian Space Science Data Centre), ISRO (Indian Space Research Organisation), for enabling user access to this dataset.
We are grateful to the LAXPC team for providing the data and requisite software tools for the analysis. This work is supported by ISRO under the \emph{AstroSat} Archival Data Utilization Program (DS 2B-13013(2)/4/2020\-Sec.2).  We thank the anonymous referee for their comments and suggestions which has improved this manuscript. We are grateful to Prof. H.M. Antia and Dr. Sunil Chandra for their valuable suggestions that helped us to address some data-related challenges.

\section{DATA AVAILABILITY}
Data analyzed in this work is publicly available on the Indian Space Science Data Center (ISSDC) website (https://astrobrowse.
issdc.gov.in/astro$\_$archive/archive/Home.jsp).

\bibliography{ref}

\end{document}